\documentclass[journal,twocolumn]{IEEEtran}
\usepackage[T1]{fontenc}
\usepackage[latin1]{inputenc}
\usepackage{color}
\usepackage{psfrag}
\usepackage[dvips]{graphicx}
\usepackage{tikz}
\usepackage{pgfplots}
\pgfplotsset{compat=newest,compat/show suggested version=false}
\usepackage{enumerate}
\usepackage{rotating}
\usepackage{srcltx}
\usepackage{booktabs}
\usepackage{multicol}
\usepackage{enumitem}
\usepackage{textcomp}
\usetikzlibrary{shapes,arrows}
\usetikzlibrary{arrows,decorations.pathmorphing,backgrounds,positioning,fit,matrix}

\usepackage{graphicx}
\usepackage{caption}
\usepackage{subcaption}

\usepackage[noadjust]{cite}
\usepackage[cmex10]{amsmath}
\usepackage{amsfonts}
\usepackage{amssymb}
\usepackage{ntheorem}
\usepackage{dsfont}    %mathds offers better set-symbols for N,Z,Q,R,C
\usepackage{mathtools} %\coloneqq for perfect :=
\usepackage{stackengine}

\definecolor{PU_orange}{HTML}{EE7F2D}
\definecolor{PU_darkorange}{HTML}{994400}
\definecolor{PU_lightorange}{HTML}{FFAA66}
\definecolor{PU_black}{HTML}{000000}
\definecolor{PU_darkgray}{HTML}{7F7F83}
\definecolor{PU_lightgray}{HTML}{BDBEC1}

\usepackage{algorithmic}
\usepackage{array}
\theoremseparator{.}
\newtheorem{corollary}{Corollary}

\newtheorem{lem}{Lemma}
\newtheorem{theorem}{Theorem}
\theorembodyfont{\upshape}
\theoremseparator{.}
\newtheorem{definition}{Definition}
\newtheorem{remark}{Remark}
\newtheorem*{example*}{Example}

\newcommand{\eu}{\mathrm{e}}

\newcommand{\mmse}{\mathrm{mmse}}

\newcommand{\E}{\mathbb{E}}

\allowdisplaybreaks

\long\def\symbolfootnote[#1]#2{\begingroup%
	\def\thefootnote{\fnsymbol{footnote}}\footnote[#1]{#2}\endgroup} %unnumbered footnote for thanks
\title{Estimation in Poisson Noise: Properties of the Conditional Mean Estimator}% 

\author{
  Alex~Dytso,~\IEEEmembership{Member,~IEEE,} and
  H.~Vincent~Poor,~\IEEEmembership{Fellow,~IEEE}% <-this % stops a space
  \thanks{This work was supported in part by the U.S. National Science Foundation under Grants CCF-0939370, CCF-1513915 and CCF-1908308. This paper was presented in part at the 2019 IEEE Information Theory Workshop \cite{ITW2019Dytso}.}
  \thanks{A.~Dytso and H.~V.~Poor are with the Department of Electrical Engineering, Princeton University, Princeton, NJ 08544, USA. E-mail: adytso@princeton.edu, poor@princeton.edu.}% <-this % stops a space
}

\begin{document}
\maketitle

\begin{abstract}
This paper considers estimation of a random variable in Poisson noise with signal scaling coefficient and dark current  as explicit parameters of the noise model.  
Specifically, the paper focuses on properties of the conditional mean estimator as a function of the scaling coefficient, the dark current parameter, the distribution of the input random variable and channel realizations. 
With respect to the scaling coefficient and the dark current,  several identities in terms of derivatives are established. For example, it is shown that the gradient  of the conditional mean estimator with respect to the scaling coefficient and dark current parameter  is proportional to the conditional variance.  Moreover, a  score function is proposed  and a Tweedie-like formula for the conditional expectation is recovered. 
With respect to the distribution, several regularity conditions are shown. For instance, it is shown that the conditional mean estimator uniquely determines the input distribution. 
Moreover, it is shown that if the conditional expectation is close to a linear function in terms of mean squared error,  then the input distribution is approximately gamma in the L\'evy distance.   Furthermore, sufficient and necessary conditions for linearity are found. Interestingly, it is shown that the conditional mean estimator cannot be linear when the dark current parameter of the Poisson noise is non-zero.
 \end{abstract}

\section{Introduction}

Poisson noise models comprise  an important set of models with a wide range of applications. 
This paper considers  denoising  of  a non-negative random  variable in \emph{Poisson noise}  with a specific focus on properties of \emph{the conditional mean estimator}.  
Concretely, for an input random variable $X \ge 0 $ the Poisson noise channel is specified by the following conditional probability mass function (pmf) of the output random variable $Y$:
\begin{align}
P_{Y|X}(y|x) = \frac{1}{y!} (ax+\lambda)^y \eu^{- (ax+\lambda)}, x\ge 0, \, y=0,1, \ldots \label{eq:PoissonTransformation}
\end{align}
where  $a>0$ is a \emph{scaling factor} and $\lambda \ge 0$ is a non-negative constant called the \emph{dark current parameter}. In words, conditioned on a non-negative input $X=x$, the output of the Poisson channel is  a non-negative integer-valued random variable $Y$ that is distributed according to  \eqref{eq:PoissonTransformation}.   In \eqref{eq:PoissonTransformation}  we use the convention that  $0^0=1$.  

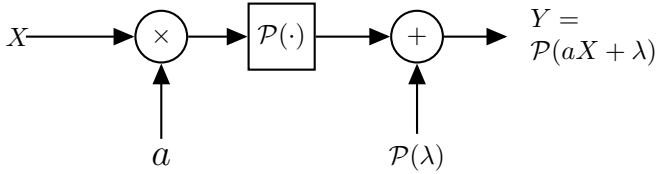
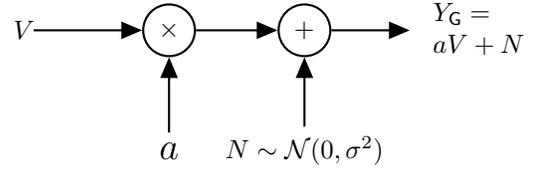
\begin{figure*}[h!]
\begin{subfigure}[a]{0.45\textwidth}
\center
%\input{FIG/DiagramPoissonChannel.tex}
%\documentclass[tikz,14pt,border=10pt]{standalone}
%\usepackage{textcomp}
%\usetikzlibrary{shapes,arrows}
%\begin{document}
% Definition of blocks:
\tikzset{%
  block/.style    = {draw, thick, rectangle, minimum height = 2.5em,
    minimum width = 2.5em},
  sum/.style      = {draw, circle, node distance = 1.8cm}, % Adder
  input/.style    = {coordinate}, % Input
  output/.style   = {coordinate} % Output
}
% Defining string as labels of certain blocks.
\newcommand{\suma}{\Large$+$}
\newcommand{\inte}{$\displaystyle \int$}
\newcommand{\derv}{\huge$\frac{d}{dt}$}

\begin{tikzpicture}[auto, thick, node distance=1.6cm, >=triangle 45]
width=6.953cm,
height=5.7cm,
\draw
	% Drawing the blocks of first filter :
	node at (-0.4,0)[right=0mm,node distance = 1cm]{$X$ }
	node [input, name=input1] {} 
	node [sum, right of=input1] (suma1) {$\times$}
	node [block, right of=suma1] (inte1) { $\mathcal{P}(\cdot)$}
         node[sum, right of=inte1] (Q1) {$+$}
         node [ below of=suma1] (ret1) {\Large$a$}
         node [ below of=Q1] (ret2) { $\mathcal{P}(\lambda)$}
         node  [ right=6.4cm](ret3){  $ \begin{array}{l} Y=\\ \mathcal{P}(aX+\lambda) \end{array}$ }
          %  node [ right of=ret3] (ret4){   }
         ;
         
%    % Joining blocks. 
%    % Commands \draw with options like [->] must be written individually
	\draw[->](input1) -- node {}(suma1);
 	\draw[->](suma1) -- node {} (inte1);
	\draw[->](inte1) -- node {} (Q1);
	\draw[->](ret1) --node {} (suma1);
	 \draw[->](ret2) --node {} (Q1);
	  \draw[->](Q1) --node {} (ret3);
\end{tikzpicture}
\caption{The Poisson noise channel with  the input $X$, the scaling factor $a$ and the dark current parameter $\lambda$.}
\label{fig:PoissonNoiseChannel}
   \end{subfigure}%
~
\hspace{1cm} 
   \begin{subfigure}[c]{0.45\textwidth}
\center
%\documentclass[tikz,14pt,border=10pt]{standalone}
%\usepackage{textcomp}
%\usetikzlibrary{shapes,arrows}
%\begin{document}
% Definition of blocks:
\tikzset{%
  block/.style    = {draw, thick, rectangle, minimum height = 3em,
    minimum width = 3em},
  sum/.style      = {draw, circle, node distance = 1.8cm}, % Adder
  input/.style    = {coordinate}, % Input
  output/.style   = {coordinate} % Output
}
% Defining string as labels of certain blocks.
\newcommand{\suma}{\Large$+$}
\newcommand{\inte}{$\displaystyle \int$}
\newcommand{\derv}{\huge$\frac{d}{dt}$}

%\begin{tikzpicture}[auto, thick, node distance=1.6cm, >=triangle 45]
%width=6.953cm,
%height=5.7cm,
%\draw
%	% Drawing the blocks of first filter :
%	node at (-0.4,0)[right=0mm,node distance = 1cm]{$X$ }
%	node [input, name=input1] {} 
%	node [sum, right of=input1] (suma1) {$\times$}
%	node [block, right of=suma1] (inte1) { $\mathcal{P}(\cdot)$}
%         node[sum, right of=inte1] (Q1) {$+$}
%         node [ below of=suma1] (ret1) {\Large$a$}
%         node [ below of=Q1] (ret2) { $\mathcal{P}(\lambda)$}
%         node  [ right=6.4cm](ret3){  $ \begin{array}{l} Y=\\ \mathcal{P}(aX+\lambda) \end{array}$ }
%          %  node [ right of=ret3] (ret4){   }
%         ;
%         
%%    % Joining blocks. 
%%    % Commands \draw with options like [->] must be written individually
%	\draw[->](input1) -- node {}(suma1);
% 	\draw[->](suma1) -- node {} (inte1);
%	\draw[->](inte1) -- node {} (Q1);
%	\draw[->](ret1) --node {} (suma1);
%	 \draw[->](ret2) --node {} (Q1);
%	  \draw[->](Q1) --node {} (ret3);
%%	% Adder
%
%\end{tikzpicture}
%\end{document}
%

%%%% Gaussian 

\begin{tikzpicture}[auto, thick, node distance=1.6cm, >=triangle 45]
width=6.953cm,
height=5.7cm,
\draw
	% Drawing the blocks of first filter :
	node at (-0.4,0)[right=0mm,node distance = 1cm]{$V$ }
	node [input, name=input1] {} 
	node [sum, right of=input1] (suma1) {$\times$}
	%node [block, right of=suma1] (inte1) { $\mathcal{P}(\cdot)$}
         node[sum, right of=suma1] (Q1) {$+$}
         node [ below of=suma1] (ret1) {\Large$a$}
         node [ below of=Q1] (ret2) { $N \sim \mathcal{N}(0,\sigma^2)$}
         node  [ right=5cm](ret3){  $ \begin{array}{l} Y_\mathsf{G}=\\ aV+N \end{array}$ }
          %  node [ right of=ret3] (ret4){   }
         ;
         
%    % Joining blocks. 
%    % Commands \draw with options like [->] must be written individually
	\draw[->](input1) -- node {}(suma1);
 	\draw[->](suma1) -- node {} (Q1);
	%\draw[->](inte1) -- node {} (Q1);
	\draw[->](ret1) --node {} (suma1);
         \draw[->](ret2) --node {} (Q1);
	  \draw[->](Q1) --node {} (ret3);
%	% Adder

\end{tikzpicture}

\caption{The Gaussian noise channel with  the input $X$, the scaling factor $a$ and the noise variance $\sigma^2$.}
\label{fig:GaussianNoiseChannel}
  \end{subfigure}%
  \caption{Noise models considered in this work.}
\end{figure*}

The random transformation of the input random variable $X$ to an output random variable $Y$  by  the channel in \eqref{eq:PoissonTransformation} will be denoted by
\begin{align}
Y=  \mathcal{P}(aX+\lambda). \label{Eq:PoissonTransformationDefinition}
\end{align}
The transformation in \eqref{Eq:PoissonTransformationDefinition} is depicted in Fig.~\ref{fig:PoissonNoiseChannel}.  It is important to note that the  operator $\mathcal{P}(\cdot)$ is not linear, and it is not true that $ \mathcal{P}(aX+\lambda)= a\mathcal{P}(X)+\lambda$.  Using the language of laser communications, the $aX$ term represents the intensity of a laser beam at the transmitter and $Y$ represents the number of photons that arrive at the receiver equipped with a particle counter (i.e., a  photodetector). The dark current parameter $\lambda$ represents the intensity of an additional source of noise or interference, which produces  an extra $\mathcal{P}(\lambda)$ photons   at a particle counter \cite{gordon1962quantum,pierce1981capacity, verdu1990asymptotic,shamai1990capacity}.  

In this work, we are interested in the properties of the conditional mean estimator of the input $X$ given the output of the Poisson channel $Y$, that is 
\begin{align}
\E[X|Y=y] = \int  x \, {\rm d} P_{X|Y=y}(x), \, y=0,1, \ldots 
\end{align}   Specifically, we are interested in how  $\E[X|Y]$ behaves as a function of the channel parameters  $(a,\lambda)$ and the distribution of $X$.   We will also study the conditional mean estimator as a function of channel realizations, that is $y \to \E[X|Y=y]$. 
  Properties of the conditional expectation are important in view of the fact that it is the unique optimal estimator under a very large family of loss functions, namely Bregman divergences \cite{banerjee2005optimality}. For example, an important member of the Bregman  family is the ubiquitous squared error loss. 

Throughout the paper, we will contrast our results with those for the \emph{Gaussian noise channel}  given by\footnote{Unlike the Poisson channel, Gaussian channel can be parameterized by only one parameter either $a$ or $\sigma^2$. However, we keep both $a$ and $\sigma^2$ for comparison.} 
\begin{align}
Y_\mathsf{G}= aV+ N,  \label{eq:GaussianChannel}
\end{align}
where $N$ is a normal random variable with zero mean and variance $\sigma^2$, $a$ is a fixed scalar,  the input $V$ is a real-valued random variable and $N$ and $V$ are independent. The transformation in \eqref{eq:GaussianChannel} is shown in Fig.~\ref{fig:GaussianNoiseChannel}. Since the behavior of the conditional expectation  $\E[V|Y_\mathsf{G}]$ is well understood, the comparison between these two channels can be very illuminating.  Also, somewhat surprisingly, we use the insights developed for the Poisson noise channel to derive a new identity for the Gaussian noise case.

The literature on the Poisson distribution is vast, and the interested reader is referred to  \cite{verdu1999poisson} and  \cite{lapidoth2009capacity} for a summary of communication theoretic applications;  \cite{CompresseSensingPoisson,wang2015signal} and \cite{wang2013designed} for applications of the Poisson model in compressed sensing;  and \cite{grandell1997mixed} and \cite{StochApproxPoissonWang} for applications of the Poisson distributions in signal processing and other fields.  Our interest in studying properties of  the conditional expectation is motivated by the bridge that the Poisson noise model offers between estimation theory and information theory. 
In \cite{guo2008mutualPoisson} and \cite{atar2012mutual} the authors have shown that information measures such as mutual information and relative entropy can be expressed as integrals over the dark current $\lambda$ and/or the scaling parameter $a$  of  Bayesian risks that use  loss functions natural (e.g., Bregman divergences) for the Poisson channel.  The results in \cite{guo2008mutualPoisson} and \cite{atar2012mutual} are counterparts of the Gaussian noise identities  between information and estimation measures shown in \cite{I-MMSE} and \cite{MismatchedMSE}.     In \cite{wang2014bregman}, the authors have generalized the results of \cite{guo2008mutualPoisson} and \cite{atar2012mutual} to the vector Poisson model and have introduced a notion of matrix Bregman divergence.  For a unifying treatment of such identities, which generalizes these results beyond the Poisson and Gaussian models, the interested reader is referred to \cite{palomar2007representation}.   Finally, for the point-wise generalizations, we refer the reader to \cite{jiao2018mutual}.

\subsection{Notation} 
Throughout the paper,  deterministic  quantities are denoted by lowercase letters and  random variables are denoted by uppercase letters.   We denote the distribution of a random variable  $X$ by $P_X$.
The expected value  and variance of $X$ are denoted by $\E[X]$ and $ \mathbb{V}(X)$, respectively.  

The binomial coefficient is denoted by 
\begin{align}
{ x \choose y} = \frac{\Gamma(x+1)}{\Gamma(y+1) \Gamma( x-y+1 )}, \, x\in \mathbb{R}, y \in \mathbb{R},
\end{align}
where $\Gamma(\cdot)$ is the gamma function. In this paper, the gamma distribution has a probability density function (pdf) given by 
\begin{align}
f(x)= \frac{\alpha^\theta}{ \Gamma(\theta)  }  x^{\theta-1} \eu^{-\alpha x},\, x \ge 0, \label{eq:pdfGamma}
\end{align}
 where $\theta>0$ is the shape parameter and $\alpha>0$ is the rate parameter.  Moreover, the mean and variance of this distribution are given by 
\begin{align}
\E[X ]=\frac{\theta}{\alpha},\, \mathbb{V}(X)=\frac{\theta}{\alpha^2}. 
\end{align}
 We denote the distribution with the pdf in \eqref{eq:pdfGamma} by $\mathsf{Gam}(\alpha, \theta)$.  
 
 \subsection{Contributions and Outline} 
 The paper outline and contributions are as follows:
 \begin{enumerate}[leftmargin=*]
 \item  In Section~\ref{sec:PropOfPoissonTrnasf} we study properties of the output distribution $P_Y$ and show: 
 
 \begin{itemize}[leftmargin=*]
 \item In Section~\ref{sec:OutputAndLaplaceTransform},  Theorem~\ref{thm:LaplaceConnection} connects the distribution  of the output $Y$ to the Laplace transform of the input $X$;
 \item In Section~\ref{sec:FunctionalPropertiesOfOutput}, Theorem~\ref{thm:propOutputDist} shows that $P_Y$ is  a continuous   and bijective  operator of the  input distribution $P_X$; and
 \item  In Section~\ref{sec:AnalyticPropertiesOutput}, studies analytical properties of $P_Y$.  In particular, 
Lemma~\ref{thm:DerivativesOuputPMF} characterizes derivatives of $P_Y$ with respect to the scaling parameter $a$ and the dark current parameter $\lambda$ and connects these derivatives to the forward and backward difference operators of $P_Y(y)$ in terms of $y$.  Moreover, Lemma~\ref{thm:TailBounds} establishes lower and upper bounds on the tail of $P_Y$. 
 \end{itemize}
 
 \item  In Section~\ref{sec:CE} we study properties of the conditional expectation of the  input $X$ given the output $Y$ and show:
 \begin{itemize}[leftmargin=*]
 \item  Section~\ref{sec:TGRformual}, Lemma~\ref{lem:FormOFConditionalExpecttion}, re-derives the known Turing-Good-Robbins Formula for the conditional expectation with explicit emphasis on the parameters $a$ and $\lambda$.  The generalizations of this formula to higher conditional moments are discussed.  Moreover,  in Lemma~\ref{lem:HigherMoments}, it is shown that any higher order conditional moment can be completely  determined by the  first order conditional moment. Furthermore, in Lemma~\ref{lem:LaplaceTransfromRepresenation} it is shown that the conditional expectation can be represented as a ratio of derivatives  of the Laplace transform of $X$.  Finally,  the Laplace representation is used  to compute conditional expectations for a few important distributions;
  \item   Section~\ref{sec:Tweedie} discusses connections between the conditional expectation and  the likelihood function, and the discrete versions of the score function.  In particular, Theorem~\ref{thm:Tweedie'sFormulaPoisson} proposes a version of the score function where instead of taking the derivative with respect to the output $y$ the gradient is taken with respect to the channel parameters $(a,\lambda)$, and the proposed score function is shown to be related to the conditional expectation via a Tweedie-like formula.  
Moreover, this new version of the score functions is compared with other definitions of score functions  known in the literature. Section~\ref{sec:Tweedie} is concluded by proposing a  version of the Fisher information and  a Brown-like identity is  establishing between the Fisher information and the minimum mean squared error. 
  
 \item   Section~\ref{sec:AnaltyticalPropertiesOfConditionalExpecatation}  studies  analytical properties of the conditional expectation.   In particular, in Theorem~\ref{thm:MonotonicityConditionalExpecation},  it is shown that the conditional expectation is a strictly increasing function of channel realizations.    Theorem~\ref{thm:MonotonictyOfConditionalExp} finds the derivative of the conditional expectation with respect to the dark current parameter $\lambda$, which is given by  the negative conditional covariance of $X$ given $Y$.  This incidentally shows that the conditional expectation is a monotonically decreasing function of $\lambda$.   Moreover, Theorem~\ref{thm:MonotonictyOfConditionalExp} finds derivatives of higher order conditional moments.    Furthermore, Corollary~\ref{cor:LimitAndMOnotonicity} finds the limiting behavior of the conditional expectation as $\lambda \to \infty$.   Finally, Lemma~\ref{lem:reverseJensen}  concludes Section~\ref{sec:AnaltyticalPropertiesOfConditionalExpecatation}  by presenting an inequality on the conditional expectation that has a flavor of the reverse Jensen's inequality;
   \item Section~\ref{sec:UniquenessOfConditionalExpecatation}  studies whether the conditional expectation is uniquely determined by the input distribution of $X$.  First, in Theorem~\ref{thm:DeterminanceOfthConditional}, as an ancillary result, it is shown that the conditional distribution of  the input $X$ given the output $Y$ is completely determined by its moments.  Second, in Theorem~\ref{thm:UniquenssOfConditionalExpectation}, it is shown the conditional expectation  uniquely determines the distribution of $X$.   The section is concluded by discussing some consequences of the uniqueness  result; and
   
   \item Section~\ref{sec:BoundsOnTheConditionalExpectation} studies upper bounds on the  conditional expectation.  In particular, in Theorem~\ref{thm:AssimptoticUppoerBoun}, it is shown that the conditional expectation is either a  linear or a sub-linear function  of the output realization $y$.  
  \end{itemize}
  
\item   In Section~\ref{sec:LinearityProperties} we  study connections between the conditional mean estimator and linear estimators and show:
\begin{itemize}[leftmargin=*]
\item  Section~\ref{sec:LinearBoundAttained}, Theorem~\ref{thm:LinearityConditions},  shows that the conditional expectation is linear if and only if the input of $X$ is according to a gamma distribution and the dark current parameter is equal to zero; and
\item Section~\ref{sec:quantitativeRefinementOfLinearity}, Theorem~\ref{thm:LevyDistanceAndLInearity}, provides a quantitative refinement of the linear condition in Theorem~\ref{thm:LinearityConditions} and shows that if the conditional expectation is close to a linear function in the $L_2$ metric   then the input distribution is close to a gamma distribution in the L\'evy metric.  
\end{itemize}
 
 \end{enumerate}

\section{Properties of the Poisson Transformation} 
\label{sec:PropOfPoissonTrnasf}

\begin{figure}[h!]  %[AD] Comment out for now
	\centering   
	%\input{FIG/BinaryOutputPoisson.tex}\label{fig:PoissonOutputDistrBinary}}%

\begin{tikzpicture}

\begin{axis}[%
width=7.5cm,
height=5.7cm,
at={(1.159in,0.77in)},
scale only axis,
xmin=0,
xmax=40,
xlabel style={font=\color{white!15!black}},
xlabel={ $y$},
ymin=0,
ymax=0.07,
ylabel style={font=\color{white!15!black}},
ylabel={$P_Y$},
axis background/.style={fill=white},
xmajorgrids,
ymajorgrids,
legend style={legend cell align=left, align=left, draw=white!15!black}
]
\addplot[ycomb, color=black, dashed, mark=o, mark options={solid, black},forget plot] table[row sep=crcr] {%
0	0.000743704427622211\\
1	0.0044630143119563\\
2	0.0133953449056532\\
3	0.0268243003168225\\
4	0.0403708924972984\\
5	0.0488752854673648\\
6	0.0500225240556494\\
7	0.0454987088209216\\
8	0.0393685508764211\\
9	0.0355692903811338\\
10	0.0362593179036452\\
11	0.0414763460201191\\
12	0.0496695546827188\\
13	0.0585322642199124\\
14	0.065779946798648\\
15	0.0697196488041524\\
16	0.0695525383862503\\
17	0.065402232451329\\
18	0.0581156576853497\\
19	0.0489332927721371\\
20	0.0391447716867362\\
21	0.0298240558000923\\
22	0.0216900772677835\\
23	0.0150887115430797\\
24	0.0100591315635709\\
25	0.00643784192904962\\
26	0.00396174835519151\\
27	0.00234770261250824\\
28	0.00134154432504168\\
29	0.000740162381065126\\
30	0.000394753268868448\\
31	0.000203743622441853\\
32	0.00010187181118344\\
33	4.93923932942463e-05\\
34	2.32434791960896e-05\\
35	1.06255904894348e-05\\
36	4.72248466193665e-06\\
37	2.04215552948055e-06\\
38	8.59854959780402e-07\\
39	3.52761009140543e-07\\
};  \addlegendentry{$\lambda=0$}

\addplot[forget plot, color=white!15!black, dashed] table[row sep=crcr] {%
0	0\\
40	0\\
};

\addplot [color=black, dashed]
  table[row sep=crcr]{%
0	0.000743704427622211\\
1	0.0044630143119563\\
2	0.0133953449056532\\
3	0.0268243003168225\\
4	0.0403708924972984\\
5	0.0488752854673648\\
6	0.0500225240556494\\
7	0.0454987088209216\\
8	0.0393685508764211\\
9	0.0355692903811338\\
10	0.0362593179036452\\
11	0.0414763460201191\\
12	0.0496695546827188\\
13	0.0585322642199124\\
14	0.065779946798648\\
15	0.0697196488041524\\
16	0.0695525383862503\\
17	0.065402232451329\\
18	0.0581156576853497\\
19	0.0489332927721371\\
20	0.0391447716867362\\
21	0.0298240558000923\\
22	0.0216900772677835\\
23	0.0150887115430797\\
24	0.0100591315635709\\
25	0.00643784192904962\\
26	0.00396174835519151\\
27	0.00234770261250824\\
28	0.00134154432504168\\
29	0.000740162381065126\\
30	0.000394753268868448\\
31	0.000203743622441853\\
32	0.00010187181118344\\
33	4.93923932942463e-05\\
34	2.32434791960896e-05\\
35	1.06255904894348e-05\\
36	4.72248466193665e-06\\
37	2.04215552948055e-06\\
38	8.59854959780402e-07\\
39	3.52761009140543e-07\\
};\addlegendentry{$\lambda=3$}

\addplot[ycomb, color=black, mark=o, mark options={solid, black},forget plot] table[row sep=crcr] {%
0	3.70268631835101e-05\\
1	0.000333280988226654\\
2	0.00150013703298304\\
3	0.00450277081004873\\
4	0.0101424429503328\\
5	0.018298990095947\\
6	0.0275833622975223\\
7	0.0358304180851622\\
8	0.0411786962824194\\
9	0.0430142565930445\\
10	0.0422003955239277\\
11	0.0405515715389922\\
12	0.0399516393895301\\
13	0.0415989237289388\\
14	0.0456608675918905\\
15	0.0513602101827264\\
16	0.0573469996605492\\
17	0.0621649261274993\\
18	0.0646541434468489\\
19	0.0641973272789751\\
20	0.0607818936394831\\
21	0.0549050416033471\\
22	0.047381949459776\\
23	0.0391275074170941\\
24	0.0309706547445416\\
25	0.0235357937000147\\
26	0.0171985748133749\\
27	0.0121024811131544\\
28	0.00821232728631965\\
29	0.00538046837700147\\
30	0.00340762339789027\\
31	0.00208854146425868\\
32	0.00124007095759808\\
33	0.000713980101912546\\
34	0.000398988841727202\\
35	0.000216593932686765\\
36	0.000114313461982284\\
37	5.87015068984279e-05\\
38	2.93507533056903e-05\\
39	1.4299084910677e-05\\
};  \addlegendentry{aa}

\addplot[forget plot, color=white!15!black] table[row sep=crcr] {%
0	0\\
40	0\\
};

\addplot [color=black]
  table[row sep=crcr]{%
0	3.70268631835101e-05\\
1	0.000333280988226654\\
2	0.00150013703298304\\
3	0.00450277081004873\\
4	0.0101424429503328\\
5	0.018298990095947\\
6	0.0275833622975223\\
7	0.0358304180851622\\
8	0.0411786962824194\\
9	0.0430142565930445\\
10	0.0422003955239277\\
11	0.0405515715389922\\
12	0.0399516393895301\\
13	0.0415989237289388\\
14	0.0456608675918905\\
15	0.0513602101827264\\
16	0.0573469996605492\\
17	0.0621649261274993\\
18	0.0646541434468489\\
19	0.0641973272789751\\
20	0.0607818936394831\\
21	0.0549050416033471\\
22	0.047381949459776\\
23	0.0391275074170941\\
24	0.0309706547445416\\
25	0.0235357937000147\\
26	0.0171985748133749\\
27	0.0121024811131544\\
28	0.00821232728631965\\
29	0.00538046837700147\\
30	0.00340762339789027\\
31	0.00208854146425868\\
32	0.00124007095759808\\
33	0.000713980101912546\\
34	0.000398988841727202\\
35	0.000216593932686765\\
36	0.000114313461982284\\
37	5.87015068984279e-05\\
38	2.93507533056903e-05\\
39	1.4299084910677e-05\\
};  \addlegendentry{$\lambda=3$}

\end{axis}
\end{tikzpicture}
\caption{Examples of the output distribution $P_Y$ for the input  $(P_X(6), P_X(16))=( 0.3,0.7 )$ with $a=1$.  The circles are the actual values of $P_Y$.  }
	\label{fig:PoissonOutput}
\end{figure}
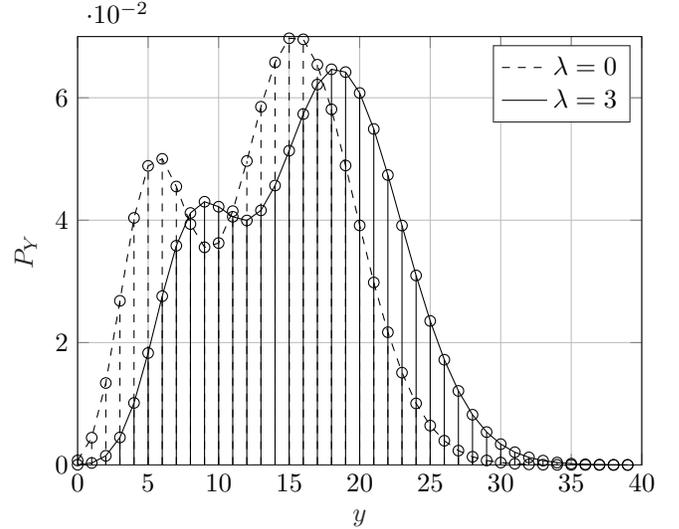

The  distribution of the output random variable $Y=\mathcal{P}(aX+\lambda)$ induced by  the input $X \sim P_X$  will be denoted by
\begin{align}
P_Y(y; P_X)=\E[ P_{Y|X}(y|X)], \,  y=0,1,\ldots  
\end{align} 
Also, for simplicity, we will use  $P_Y(y )$ instead of  $P_Y(y; P_X)$ whenever the nature of the   underlying input probability  distribution is clear or nonessential. Examples of the output distribution induced by the binary input are shown in Fig.~\ref{fig:PoissonOutput}.

\subsection{Connections to the Laplace Transform} 
\label{sec:OutputAndLaplaceTransform}
This section shows an alternative representation of  $P_Y$ in terms of the Laplace transform of $X$. 

\begin{theorem}\label{thm:LaplaceConnection}
Denote the Laplace transform  of a non-negative random variable $W$ by 
\begin{align}
\mathcal{L}_W(t) = \E \left[ \eu^{-tW} \right], t \ge 0, 
\end{align}
 and its $n$-th derivative by  $\mathcal{L}_W^{(n)}(t)$.  Then,  for every $a>0, \lambda \ge 0$  and  $y=0,1,\ldots$ 
\begin{align}
P_{Y}(y;P_{X}) =  \frac{(-1)^y \eu^{-\lambda}  }{y!}     \sum_{i=0}^y { y \choose i} a^{y-i} (- \lambda)^{i}
  \mathcal{L}_{X}^{(y-i)}(a),  \label{eq:LaplaceAndOutput}
\end{align} 
and
\begin{align}
\mathcal{L}_{aX+\lambda}(t)= \sum_{y=0}^\infty   (-1)^y P_{Y}(y;P_X) (t-1)^y, \, \forall  \, |t-1| < 1.  \label{eq:ConnectionToLaplaceTransform}
\end{align}
\end{theorem}
\begin{IEEEproof}
See Appendix~\ref{app:thm:LaplaceConnection}.
\end{IEEEproof}

For $\lambda=0$ the expression in \eqref{eq:LaplaceAndOutput} takes  a simple form  given by 
\begin{align}
P_{Y}(y;P_{X}) =  \frac{(-a)^y}{y!}   
  \mathcal{L}_{X}^{(y)}(a), \, y=0,1,\ldots  \label{eq:LaplaceConnectionZeroDarkCurrent}
\end{align}
To show the utility of the expression in \eqref{eq:LaplaceAndOutput}, consider an input $X\sim \mathsf{Gam}(\alpha, \theta)$ with the Laplace transform 
\begin{align}
\mathcal{L}_X(t)=  \frac{1}{(1+\frac{ t}{\alpha})^\theta}, \, t > -\alpha,
\end{align}
and the channel output given  by $Y=\mathcal{P}(aX)$. 
Using  \eqref{eq:LaplaceAndOutput}  we arrive at the following output distribution: 
\begin{align}
P_{Y}(y)&= (-a)^y   \frac{ \alpha^\theta}{  \left(\alpha+ a\right)^{\theta+y} } { {-\theta} \choose y} \\
&=  \frac{ a^y  \alpha^\theta}{  \left(\alpha+ a\right)^{\theta+y} } { {\theta+y-1} \choose y},\, y=0,1,\ldots \label{eq:NegativeBinomial}
\end{align}
The distribution in \eqref{eq:NegativeBinomial} is known as the negative binomial distribution with failure parameter $\theta$ and success probability  $\frac{a}{\alpha+a}$. 

\begin{remark}\label{rem:WeiestrssTransform}  There exists a similar result to Theorem~\ref{thm:LaplaceConnection}  for the Gaussian noise channel in \eqref{eq:GaussianChannel}. Specifically, the probability density function (pdf) of the output $Y_\mathsf{G}$  is a  \emph{Weierstrass transform} of the input $V$ \cite{zayed1996handbook}. 
\end{remark}

\subsection{Properties of the Output Distribution as a Function of the Input Distribution} 
\label{sec:FunctionalPropertiesOfOutput}
In this section, we are also interested in how  $P_Y$ behaves as a function of $P_X$.  We will need the following definition. 

\begin{definition}\label{def:WeakConvergence}
	A sequence of probability measures $\{P_{X_n}\}_{n\in\mathbb{N}}$ on $\mathbb{R}$ is said to \emph{converge weakly} to the probability measure $P_X$   (i.e., $X_n \to X$ \emph{in distribution})  if  and only if
	\begin{align}
		\lim_{n \to \infty} \E\left[ \phi(X_n) \right] =  \E\left[ \phi(X)\right], 
	\end{align}%
for all bounded and continuous functions $\phi$.
\end{definition}%

The next result presents two important properties of the output distribution. 
\begin{theorem}\label{thm:propOutputDist}  Let $Y= \mathcal{P}(aX+\lambda)$. Then, for all $a>0$ and $\lambda \ge 0$,  $P_Y$  satisfies the following properties: 
\begin{itemize}[leftmargin=*]
\item Let $P_{X_n} \to P_X$ weakly.  Then, 
 $P_{Y_n}(\cdot;P_{X_n}) \to  P_{Y}(\cdot; P_{X})$ weakly. In other words, the mapping $X \to \mathcal{P}(aX+\lambda)$ is continuous in distribution;  and 
\item $P_Y$ is a bijective operator of $P_X$, that is 
\begin{align}
 P_{Y_1}( \cdot ;P_{X_1})= P_{Y_2}( \cdot ; P_{X_2})   \Longleftrightarrow  P_{X_1}=P_{X_2}. 
\end{align}

\end{itemize} 
\end{theorem}
\begin{IEEEproof}
To establish continuity  observe that for every $k$ the Poisson probability $ P_{Y|X}(k|x)$  is a bounded  and continuous function of $x$ and, therefore,  by the definition of   convergence in distribution, for every non-negative integer $k$
\begin{align}
\lim_{n \to \infty} P_{Y_n}(k; P_{X_n}) &=  \lim_{n \to \infty}  \E[  P_{Y|X}(k|X_n) ] \notag\\
&=\E[  P_{Y|X}(k|X) ]=P_{Y}(k;P_{X}).  \label{eq:ConvergencePMF}
\end{align} 
 Thus, if  $P_{X_n} \to P_X$ weakly,  then  $P_{Y_n} \to P_Y$ weakly\footnote{Note that in \eqref{eq:ConvergencePMF} we have established a point-wise convergence of the pmf. However, for integer value random  variables (note our output is integer valued) point-wise convergence in the pmf is equivalent to  weak convergence \cite[Ch. 8.8, Exercise 4]{resnick2013probability}.}.   This concludes the proof of the continuity of the mapping $P_X \to P_Y(\cdot; P_X)$.

Next we show that the output distribution is a bijective operator of $P_X$.  The implication 
\begin{align}
P_{X_1}=P_{X_2}  \Longrightarrow   P_{Y_1}( \cdot; P_{X_1})= P_{Y_2}( \cdot ;P_{X_2})  
\end{align} 
is immediate. Therefore, it remains to show that $P_{Y}( \cdot; P_{X})$ is an injective operator.  

 The injectivity follows from the fact that, in view of \eqref{eq:ConnectionToLaplaceTransform}, the output pmf $ P_{Y}( \cdot; P_{X})$ completely determines the Laplace transform of $P_X$ on the interval $t \in (0,2)$, and since  the Laplace transform of $P_X$ is unique on any given  interval \cite[Sec.~30]{billingsley2008probability}, we have that $ P_{Y}(\cdot; P_{X})$ completely determines  $P_X$.   This concludes the part of the proof that shows that  $ P_{X} \to P_{Y}(\cdot; P_{X})$ is an injective transform. 

This concludes the proof of the theorem. 
 \end{IEEEproof}

 \subsection{Analytical Properties of the Output Distribution }
 \label{sec:AnalyticPropertiesOutput}
 
Another useful identity of the Poisson transformation relates the derivative of $P_Y$ with respect to the channel parameter to the forward and backward differences of $yP_Y(y)$ and $P_Y(y)$. 
  \begin{lem}\label{thm:DerivativesOuputPMF} Let $Y=\mathcal{P}(aX+\lambda)$. Then, for every $a>0,\lambda \ge 0$ and   $ y=0,1,\ldots$
\begin{align}
\frac{\rm d}{ {\rm d}  a } P_{Y|X}(y|x)&=    x  \frac{\rm d}{ {\rm d}  \lambda } P_{Y|X}(y|x)  \\
&=x \left(P_{Y|X}(y-1|x)-P_{Y|X}(y|x)\right) , \label{eq:ChannelDerivatives}
\end{align}
and 
\begin{align}
 a \frac{ \rm d}{ { \rm d} a } P_Y(y) +\lambda \frac{\rm d }{ {\rm d} \lambda  }  P_Y(y) & =  y  P_Y(y)-(y+1) P_Y(y+1),   \label{eq:GradientTypeConnection}\\
 \frac{{ \rm d} }{{ \rm d} \lambda}  P_Y(y) &=   P_Y(y-1) -   P_Y(y), \label{eq:DarkCurrentDerivative} 
 % a^2 \frac{ \rm d^2}{ { \rm d} a^2 } P_Y(y) + a \lambda   \frac{ \rm d}{ { \rm d} a }  \frac{ \rm d}{ { \rm d} \lambda  } P_Y(y) &=  (y-1)   a \frac{ \rm d}{ { \rm d} a } P_Y(y)-(y+1)   a \frac{ \rm d}{ { \rm d} a } P_Y(y+1).  
\end{align}
where  $P_{Y|X}(-1|x)=P_Y(-1)=0$. 
\end{lem} 
\begin{IEEEproof}
See Appendix~\ref{app:thm:DerivativesOuputPMF}. 
\end{IEEEproof}

Special cases of Lemma~\ref{thm:DerivativesOuputPMF} have been shown in the past; see for example \cite[Lemma~8.5]{guo2013interplay}. Lemma~\ref{thm:DerivativesOuputPMF} will be used in later sections to study properties of the conditional expectation.  For example, to study monotonicity properties of the mapping $y \to \E[X|Y=y]$ it will be convenient to translate the differences with respect to discrete points in $y$ into the derivatives with respect to the continuous parameter $\lambda$.

\begin{remark}\label{rem:RelationToDifferenceOperators} 
The identity in \eqref{eq:GradientTypeConnection} can be expressed as an equality between the  inner product  of the vector ${\bf v}=[a,\lambda]$ with the gradient of $P_Y$ with respect to ${\bf v}=[a,\lambda]$,   and  the forward difference operator as follows:
\begin{align}
 {\bf v} \boldsymbol{\cdot} \nabla_{{\bf v}} P_Y(y)  =-   \mathsf{D}^{\mathsf{fw}}_1 [ yP_Y(y) ], \label{eq:DifferenceOperator}
\end{align}
where   `$\,  \boldsymbol{\cdot} \, $' denotes the  inner product, and  the forward difference operator  of a function $f$ is given by   
\begin{align}
\mathsf{D}^{\mathsf{fw}}_k[f(y)]= f(y+k)-f(y).
\end{align}
    Similarly, the expression in \eqref{eq:DarkCurrentDerivative}, can be expressed in terms of the backward difference operator as 
\begin{align}
 \frac{{ \rm d} }{{ \rm d} \lambda}  P_Y(y)= -\mathsf{D}^{\mathsf{bw}}_1[ P_Y(y)],
 \end{align}
where  the backward difference operator  of a function $f$ is given by   
\begin{align}
 \mathsf{D}^{\mathsf{bw}}_k[f(y)]= f(y)-f(y-k).
 \end{align}
 
% Another interesting observation is that the second derivative with respect to  $\lambda$, which can recursively be derived from \eqref{eq:DarkCurrentDerivative},   is given by 
% \begin{align}
% \frac{ {\rm d}^2}{ {\rm d} \lambda^2} P_Y(y)=    P_Y(y)- P_Y(y-2)=   \mathsf{D}^{\mathsf{bw}}_2[ P_Y(y)].  \eqref{eq:HeatLikeEquation}
% \end{align} 
% In other words,  \eqref{eq:HeatLikeEquation} relates second derivative with respect to dark current to the second order 
\end{remark}

\begin{remark}  Consider a Gaussian noise channel in \eqref{eq:GaussianChannel}. Let $f_{Y_{\mathsf{G}}|V}$ be the conditional pdf for this channel.  It is not difficult to check that for this channel we have the following  identity between derivatives with respect to the realization of the output $y$ and the scaling parameter $a$:  \begin{align}
\frac{{\rm d}}{ {\rm d} a} f_{Y_{\mathsf{G}}|V}(y|v)= - \frac{v}{\sigma^2} \, \frac{{\rm d}}{ {\rm d} y} f_{Y_{\mathsf{G}}|V}(y|v).
\end{align} 
Continuing with our  parallelism between derivatives and difference operators the identity in  \eqref{eq:ChannelDerivatives} can be written as follows:
\begin{align}
 \frac{\rm d}{ {\rm d} a } P_{Y|X}(y|x) =-x \,  \mathsf{D}^{\mathsf{bw}}_1[P_{Y|X}(y|x)]. 
\end{align} 
\end{remark}

Another useful result concerns the following lower and upper bounds on the tail of $P_Y$.

\begin{lem}\label{thm:TailBounds}
 Let $Y=\mathcal{P}(aX+\lambda)$. Then, for every $a>0,\lambda \ge 0$ and  $ y=0,1,\ldots$
\begin{align}
 \frac{1}{y!}  \eu^{ y  \E[ \log(aX+\lambda)]- (a\E[X]+\lambda) } &\le  P_Y(y)  \le  \frac{1}{y!}  y^y \eu^{-y}  \le   \frac{1}{\sqrt{2 \pi \, y}} .\label{eq:TailOverAllSitributions}
\end{align}
\end{lem} 
\begin{IEEEproof}
See Appendix~\ref{app:thm:TailBounds}. 
\end{IEEEproof}

Properties of the output distribution $P_Y$ derived in this section will be useful in the next section where we study properties of the conditional expectation.

\section{Properties of the Conditional Expectation}
\label{sec:CE}

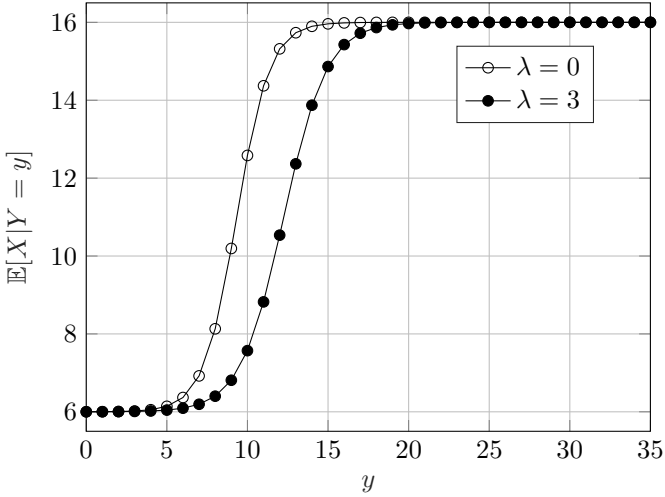
\begin{figure}[h!]  
	\centering   
\definecolor{mycolor1}{rgb}{0.00000,0.44700,0.74100}%
\definecolor{mycolor2}{rgb}{0.85000,0.32500,0.09800}%
\begin{tikzpicture}

\begin{axis}[%
width=7.5cm,
height=5.7cm,
at={(1.011in,0.642in)},
scale only axis,
xmin=0,
xmax=35,
xlabel style={font=\color{white!15!black}},
xlabel={$y$},
ymin=5.5,
ymax=16.5,
ylabel style={font=\color{white!15!black}},
ylabel={$ \E[X|Y=y]$},
axis background/.style={fill=white},
xmajorgrids,
ymajorgrids,
legend style={legend cell align=left, align=left, draw=white!15!black,at={(0.9,0.9)}}
]
\addplot [color=black,  mark=o]
  table[row sep=crcr]{%
0	6.00105921948798\\
1	6.00282408674666\\
2	6.00752735500718\\
3	6.02004779554012\\
4	6.05328275447905\\
5	6.1408366511599\\
6	6.3669510437365\\
7	6.92213944468126\\
8	8.13145534452311\\
9	10.1939953018791\\
10	12.5826913631887\\
11	14.3704716877303\\
12	15.319634728346\\
13	15.7335320520162\\
14	15.8983821507709\\
15	15.9616497396029\\
16	15.9855840990038\\
17	15.9945891619949\\
18	15.9979702493323\\
19	15.9992387469275\\
20	15.999714516515\\
21	15.9998929417829\\
22	15.9999598529\\
23	15.9999849447997\\
24	15.9999943542946\\
25	15.9999978828597\\
26	15.9999992060723\\
27	15.9999997022771\\
28	15.9999998883539\\
29	15.9999999581327\\
30	15.9999999842998\\
31	15.9999999941124\\
32	15.9999999977922\\
33	15.9999999991721\\
34	15.9999999996895\\
35	15.9999999998836\\
36	15.9999999999563\\
37	15.9999999999836\\
38	15.9999999999939\\
};
\addlegendentry{$\lambda=0$}

\addplot [color=black, mark=*, mark options={solid, black}]
  table[row sep=crcr]{%
0	6.00105921948798\\
1	6.00223586688865\\
2	6.00471899109427\\
3	6.00995709370606\\
4	6.0209973009247\\
5	6.04422445814593\\
6	6.09290621972743\\
7	6.19413135164548\\
8	6.40117935454598\\
9	6.81079271535094\\
10	7.57021578567893\\
11	8.82246838974852\\
12	10.5360154611809\\
13	12.3670357063051\\
14	13.8722845922823\\
15	14.8650358186692\\
16	15.4282307779477\\
17	15.7207587065478\\
18	15.8657548189972\\
19	15.9359576841416\\
20	15.9695615689294\\
21	15.985558660455\\
22	15.9931541621593\\
23	15.996756065894\\
24	15.9984631372403\\
25	15.9992719534871\\
26	15.9996551226472\\
27	15.9998366340781\\
28	15.9999226154769\\
29	15.999963344024\\
30	15.9999826366094\\
31	15.9999917752285\\
32	15.9999961040539\\
33	15.9999981545515\\
34	15.9999991258401\\
35	15.9999995859242\\
36	15.9999998038588\\
37	15.999999907091\\
38	15.9999999559905\\
};
\addlegendentry{$\lambda=3$}

\end{axis}
\end{tikzpicture}%
		\caption{Examples of conditional expectations for the input  $(P_X(6), P_X(16))=( 0.3,0.7 )$ with $a=1$ and dark current values $\lambda=0,3$.  The circles are the actual values of $\E[X|Y]$.  }
	\label{fig:PoissonConditionalExpectation}
\end{figure}

In this section, we study the properties of the conditional expectation. Specifically, we focus on how $\E[X|Y=y]$ behaves as a function of the channel parameters $(a,\lambda)$, the channel realization $y$, and the distribution of $X$.  Examples of conditional expectations for the binary distribution used in Fig.~\ref{fig:PoissonOutput} are shown in Fig.~\ref{fig:PoissonConditionalExpectation}. 

Next, we present a formula that will form a basis for much of our analysis of the conditional mean estimator. 
 
\subsection{The Turing-Good-Robbins Formula }
\label{sec:TGRformual}

An interesting property of the conditional expectation over a Poisson noise channel is its dependence only on the marginal distribution  of $Y$.  This identity was first demonstrated by Good in \cite{good1953population} and is  credited to Alan Turing. Moreover, it has also been independently derived by Robbins in \cite{robbins1956empirical} in the context of empirical Bayes estimation.  For completeness, we derive it next. 

\begin{lem}\label{lem:FormOFConditionalExpecttion} Let $Y=\mathcal{P}(a X+\lambda)$. Then, for every $a>0$ and $\lambda \ge 0$, 
\begin{align}
\E[X|Y=y]=  \frac{1}{a}  \frac{ (y+1) P_Y(y+1)}{P_Y(y)}-\frac{\lambda}{a}, \,y=0,1,\ldots \label{eq:COnditionalExpectation}
\end{align}
\end{lem} 
\begin{IEEEproof}
The proof follows via the following sequence of manipulations:
\begin{align}
&\E[X|Y=y]  \notag\\
&=  \frac{ \E\left[ X  P_{Y|X}(y|X)   \right]}{ P_Y(y) }\\
&=  \frac{ \E\left[ X  \frac{1}{y!}(  aX+\lambda  )^y \eu^{-(  aX+\lambda  )}    \right]}{ P_Y(y) }\\
&= \frac{1}{a }\frac{ \E\left[   \frac{1}{y!}(  aX+\lambda  )^{y+1} \eu^{-(  aX+\lambda  )}    \right]   - \lambda   P_Y(y) }{ P_Y(y) }\\
&= \frac{1}{a }\frac{ (y+1) \E\left[   \frac{1}{(y+1)!}(  aX+\lambda  )^{y+1} \eu^{-(  aX+\lambda  )}    \right]   - \lambda   P_Y(y) }{ P_Y(y) }\\
&= \frac{1}{a }\frac{ (y+1) P_Y(y+1)  - \lambda   P_Y(y) }{ P_Y(y) }.
\end{align}
This completes the proof of \eqref{eq:COnditionalExpectation}.   
\end{IEEEproof}

The key advantage of the expression in \eqref{eq:COnditionalExpectation} is that it  depends only on the marginal $P_Y$. This, in turn, avoids computation of the often complicated conditional $P_{X|Y}$.   The \emph{Turing-Good-Robbins} (TGR) formula played an important role in the development of \emph{empirical Bayes estimation}; the interested reader is referred to \cite[Chapter 6.1]{efron2016computer} for a historical account and impact of the TGR formula.    Vector versions of Lemma~\ref{lem:FormOFConditionalExpecttion} were shown in \cite[Lemma~3]{palomar2007representation}  and  \cite[Lemma~3]{wang2014bregman}.

\begin{remark} It is not difficult to see that the expression in \eqref{eq:COnditionalExpectation} can be generalized to higher order moments as follows: for every    non-negative integers $k$ and $y$
\begin{align}
\E \left[ ( aX +\lambda)^k |Y =y \right]= \frac{(y+k)!}{ y!}  \frac{P_Y(y+k)}{P_Y(y)}.  \label{eq:Hihger Moments}
\end{align} 
This, for example, leads to the following interesting expression for the conditional variance of $U=aX +\lambda$:
\begin{align}
 \hspace{-0.01cm} \mathbb{V}(U|Y \hspace{-0.01cm} = \hspace{-0.01cm} y)  
&= \E[U|Y=y] \hspace{-0.05cm} \left( \E[U |Y=y+1]- \hspace{-0.02cm} \E[U|Y \hspace{-0.02cm}=y] \right). \label{eq:COnditionalVariance}
\end{align}
\end{remark}

Combining the  TGR formula with the expression in \eqref{eq:Hihger Moments}, we arrive at the following lemma that relates all the conditional moments to the first conditional moment. 

\begin{lem}\label{lem:HigherMoments} Let $Y=\mathcal{P}(aX+\lambda)$. Then, for every positive integer $k$ and every non-negative  integer $y$ 
\begin{align}
\E \left[ ( aX +\lambda)^k |Y =y \right]= \prod_{i=0}^{k-1}  \E \left[  aX +\lambda |Y =y+i \right].  \label{eq:productIdentityForConditionalMoments}
\end{align}
\end{lem}
\begin{IEEEproof}
The proof follows by combining    \eqref{eq:COnditionalExpectation} and  \eqref{eq:Hihger Moments}. 
\end{IEEEproof}

In probability theory, typically,  lower order moments are controlled by higher order moments (i.e., Jensen's inequality).  The identity in \eqref{eq:productIdentityForConditionalMoments} is very strong and implies that all of the higher conditional moments can be recovered from the first  conditional moment.  This observation will be used in Section~\ref{sec:UniquenessOfConditionalExpecatation} to show that the conditional expectation is  unique for every input distribution.

\begin{remark}  \label{rem:GaussianHigherMoments}
To the best of our knowledge no Gaussian counterpart of the identity in \eqref{eq:productIdentityForConditionalMoments} has ever been presented in the past.   In fact, inspired by the result in Lemma~\ref{lem:HigherMoments} the authors of this paper have recently derived in \cite{dytsoPorofIntegralHiherMoments} the following Gaussian analog of the identity in \eqref{eq:productIdentityForConditionalMoments}: 
\begin{align}
&\E[(aV)^k | Y_{\mathsf{G}}=y] \notag\\
& =  \sigma^{2k} \eu^{-\frac{1}{\sigma^2} \int_0^y  \E[ a V | Y_{\mathsf{G}}=t] {\rm d} t}  \frac{{\rm d}^k}{ {\rm d} y^k }   \eu^{  \frac{1}{\sigma^2} \int_0^y  \E[ aV | Y_{\mathsf{G}}=t] {\rm d} t}, \label{eq:GaussianCounterpartForCOntionalMoments}
\end{align}
where $V$ and $Y_{\mathsf{G}}$ are related through \eqref{eq:GaussianChannel}.   Note that \eqref{eq:GaussianCounterpartForCOntionalMoments} is in the same spirit as \eqref{lem:HigherMoments} in the sense that all the higher conditional moments can be characterized by the first  conditional moment.  
\end{remark}

  It is also interesting to combine the  TGR formula with the Laplace transform representation of the output pmf in \eqref{eq:LaplaceAndOutput}, which leads to the following representation of the conditional expectation. 
  
  \begin{lem}\label{lem:LaplaceTransfromRepresenation}   Let $Y=\mathcal{P}(a X+\lambda)$. Then, for every $a>0$ and $\lambda \ge 0$
  \begin{align}
& \E[X|Y=y] \notag\\
& = \hspace{-0.1cm} -   \frac{  \sum_{i=0}^{y+1} { y+1 \choose i} a^{y+1-i} (- \lambda)^{i}
  \mathcal{L}_{X}^{(y+1-i)}(a)}{ a \sum_{i=0}^y { y \choose i} a^{y-i} (- \lambda)^{i} 
  \mathcal{L}_{X}^{(y-i)}(a)}   -\frac{\lambda}{ a}, \, y=0, 1, \ldots
  \end{align} 
  In particular, for $\lambda=0$ 
\begin{align}
\E[X|Y=y] = % \frac{1}{a}(y+1)  \frac{ \frac{(-a)^{y+1}}{(y+1)!}   
 % \mathcal{L}_{X}^{(y+1)}(a)}{  \frac{(-a)^y}{y!}   
 % \mathcal{L}_{X}^{(y)}(a) }
 - \frac{ \mathcal{L}_{X}^{(y+1)}(a)}{ \mathcal{L}_{X}^{(y)}(a)}, \, y=0, 1, \ldots  \label{eq:LaplaceRepresentation}
\end{align} 
\end{lem}
The expression in \eqref{eq:LaplaceRepresentation} enables computation of the conditional expectation if the Laplace transform of the random variable is known and can be easily  differentiated.  Using \eqref{eq:LaplaceRepresentation},  Table~\ref{table:CEexamples}  collects expressions for the conditional expectations of a few important distributions. 

\begin{table*}[h!]
\caption{Examples of Conditional Expectations for $\lambda=0$.} \label{table:CEexamples}
\center
\begin{tabular}{|p{4.7cm}|  p{3.2cm}| p{5.5cm}|   p{3.2cm}| }
\hline
Distribution  of $X$ &  Laplace Transform $\mathcal{L}_X(t)$  &   $\mathcal{L}_X^{(n)}(t)$    &     $\E[X|Y=y], \, y=0,1,\ldots $  \\
\hline   
Gamma (defined in \eqref{eq:pdfGamma})  &  $  \frac{1}{(1+\frac{ t}{\alpha})^\theta}, \,   t>-\alpha$    &    $(-1)^n  \frac{1}{\alpha^n}  \frac{1}{(1+  \frac{t}{\alpha})^{n+\theta} } \frac{\Gamma(\theta+n)}{\Gamma(\theta)},   t>-\alpha $  &  $\frac{y}{\alpha +a} +\frac{\theta}{\alpha+a} $ \\ 
\hline
Inverse-gamma  \newline $f_X(x)=\frac{\beta^\alpha}{\Gamma(\alpha)} \frac{1}{x^{\alpha+1}}  \eu^{-\frac{\beta}{x}}, x>0, \alpha>0,\beta>0$  &   $\frac{2 (\beta t)^\frac{\alpha}{2}}{\Gamma(\alpha)} K_\alpha( \sqrt{4 \beta t}), t>0$ where $K_\alpha( \cdot)$ is a modified Bessel function of a second kind     &   $(-1)^n  \frac{\beta^\alpha}{\Gamma(\alpha)}     2K_{\alpha-n}\left(\sqrt{ 4 \beta t} \right)\,{\left(\frac{\beta}{t}\right)}^{\frac{n-\alpha}{2}}, t>0$ &  $ {\left(\frac{\beta}{a}\right)}^{\frac{1}{2}}  \frac{    K_{\alpha-(y+1)}\left(\sqrt{ 4 \beta a} \right)}{    K_{\alpha-y}\left(\sqrt{ 4 \beta a} \right)}$ \\ 
\hline
Uniform \newline $f_X(x)=\frac{1}{b-c},  0\le c \le x \le b $  &  $-  \frac{\eu^{-tb}- \eu^{-tc}}{t (b-c)}$   &   $(-1)^n  \frac{1}{b-c}      \frac{\Gamma \left(n+1,c\,t\right)-\Gamma \left(n+1,b\,t\right)}{t^{n+1}}$ where $\Gamma(\cdot, \cdot)$ is the upper incomplete Gamma function &  $ \frac{        \Gamma \left(y+2,c \,a\right)-\Gamma \left(y+2,b\,a\right) }{ a  \left(   \Gamma \left(y+1,c\,a\right)-\Gamma \left(y+1,b\,a\right)  \right) } $  \\ 
\hline
Bernoulli  \newline $\mathbb{P}[X=1] =p=1-\mathbb{P}[X=0] =1-q$   &   ${\displaystyle q+p\eu^{-t}}$  &  $p (-1)^n     \eu^{-t}$  &  $  \left \{\begin{array}{ll} \frac{  p \eu^{-a}}{ q+p \eu^{-a}  }  & y=0  \\ 
1 & y > 0 
\end{array}\right.$   \\ 
\hline
Poisson  \newline $\mathbb{P}[X=k] = \frac{\gamma^k \eu^{-\gamma}}{k!}, k=0,1,\ldots $   & $ \eu^{  \gamma (\eu^{-t}-1)}$  &  $ (-1)^n  \eu^{  \gamma (\eu^{-t}-1)} T_n (  \eu^{  \gamma \eu^{-t}}  )$ where $T_n(\cdot)$ is the  Touchard  polynomial  &    $\frac{T_{y+1}( \eu^{  \gamma \eu^{-a}}  )}{ T_{y}( \eu^{  \gamma \eu^{-a}}  )}$  \\ 
\hline
Exponential Family  \newline  $\frac{ {\rm d} P_{X} }{d \mu}(x;\theta)=  \eu^{ \theta x -\psi(\theta)}, x \ge 0, \theta \in \mathcal{N} $  where
\begin{itemize}[leftmargin=*]
\item ${\rm d}\mu$ - \emph{base measure};
\item  $\theta$ - \emph{natural parameter};
\item  $\mathcal{N} \subseteq \mathbb{R}$ - \emph{natural parameter space}; and
\item  $\psi$ - \emph{log-partition function}. 
\end{itemize}
   &  $\eu^{\psi(\theta-t)-\psi(\theta)}, \,  \theta-t \in  \mathcal{N}  $  &   $ \eu^{-\psi(\theta)} \frac{ {\rm d}^n}{ {\rm d} t^n} \eu^{\psi(\theta-t)}$    &   $ - \frac{ \frac{ {\rm d}^{y+1}}{ {\rm d} t^{y+1}} \eu^{\psi(\theta-t)}}{ \frac{ {\rm d}^y}{ {\rm d} t^y} \eu^{\psi(\theta-t)}} \Big |_{t=a}$    \\ 
\hline
\end{tabular}
\end{table*}

\subsection{On Tweedie's Formula, the Score Function, the Fisher Information and Brown's Identity} 
\label{sec:Tweedie}
Consider a Gaussian channel  given  in \eqref{eq:GaussianChannel} for which the classical \emph{Tweedie's formula}  for the conditional expectation is given by 
\begin{align}
a \E[V|Y_\mathsf{G}=y]&= y + \sigma^2 \frac{\rm d}{{\rm d} y} \log (f_{Y_\mathsf{G}}(y))\\
&=  y + \sigma^2 \rho_{Y_\mathsf{G}}(y), \label{eq:TweediesFormulaGaussian}
\end{align}
where the quantity
\begin{align}
\rho_{Y_\mathsf{G}}(y)=  \frac{\rm d}{{\rm d} y} \log  f_{Y_\mathsf{G}}(y)=  \frac{f_{Y_\mathsf{G}}'(y)}{f_{Y_\mathsf{G}}(y)},
\end{align} 
is known as the \emph{score function} and the logarithm of the pdf is known as the \emph{log-likelihood} function.  The identity in \eqref{eq:TweediesFormulaGaussian} has been derived by Robbins in \cite{robbins1956empirical} where he credits  Maurice Tweedie for the derivation.  The version of \eqref{eq:TweediesFormulaGaussian} for multivariate normal has been derived by Esposito  in \cite{esposito1968relation}.  The identity in \eqref{eq:TweediesFormulaGaussian}   has an advantage that it  depends only on the marginal distribution of the output; see  \cite{dytso2019capacity}  for an application example.  In this section, we propose an analog of Tweedie's formula for the Poisson case. 

Note  that, in the Poisson case it is not possible to obtain the logarithmic derivative form similar to the one in \eqref{eq:TweediesFormulaGaussian} in view of the fact that the output space is discrete.    Nonetheless, it is interesting to observe the following forward difference property of the logarithm of $P_Y$, which is   just a restatement of the  TGR formula in \eqref{eq:COnditionalExpectation}. 
\begin{corollary} Let $Y=\mathcal{P}(a X+\lambda)$. Then,
\begin{align}
\hspace{-0.25cm} \mathsf{D}^{\mathsf{fw}}_1 \left[ \log P_Y(y )  \right]= \log\left(  \E[aX+\lambda |Y=y] \right)  -\log(y+1).  \label{eq:DifferenceOfLogFunction}
\end{align}
\end{corollary}
The identity in \eqref{eq:DifferenceOfLogFunction} has been previously demonstrated in \cite{reveillac2013likelihood}.

 Although for the Poisson case, there is no Tweedie's formula that differentiates with respect to $y$, using the result in Lemma~\ref{thm:DerivativesOuputPMF} that translates the difference with respect to discrete points to the gradient with respect to continuous parameters $a$ and $\lambda$, we propose the following version of Tweedie's formula for the Poisson noise channel. The proposed formula  does have  a logarithmic derivative form.  

\begin{theorem}\label{thm:Tweedie'sFormulaPoisson} For  $Y=\mathcal{P}(a X+\lambda)$ let ${\bf v}=[a,\lambda]$ and define a Poisson score function as
\begin{align}
\rho^{\mathsf{Po}}_Y(y)&=  - {\bf v} \boldsymbol{\cdot}  \nabla_{\bf v} \log  P_Y(y) , \label{eq:GradientDefintionoPoisson}
\end{align} 
and its scaling and dark current components, respectively,  by 
\begin{align}
\rho^{\mathsf{Po}, \mathsf{scale}}_Y(y)&= \frac{ \rm d}{ { \rm d} a } \log(   P_Y(y))=\frac{  \frac{ \rm d}{ { \rm d} a } P_Y(y)   }{P_Y(y)},\\
\rho^{\mathsf{Po}, \mathsf{d.c}}_Y(y)&=\frac{ \rm d}{ { \rm d} \lambda } \log(   P_Y(y))=\frac{ \frac{\rm d }{{\rm d} \lambda  }  P_Y(y)  }{P_Y(y)},
\end{align} 
and define  discrete  forward and backward score functions, respectively,  by 
\begin{align}
\rho^{\mathsf{Po}, \mathsf{fw.d} }_Y(y) &=  \frac{    \mathsf{D}^{\mathsf{fw}}_1\left[y P_Y(y) \right]   }{P_Y(y)},\\
\rho^{\mathsf{Po}, \mathsf{bw.d} }_Y(y) &=  \frac{    \mathsf{D}^{\mathsf{bw}}_1\left[ P_Y(y) \right]   }{P_Y(y)}. 
\end{align} 
Then,   
\begin{align}
\E \left[aX+\lambda| Y=y \right]=y + \rho^{\mathsf{Po}}_Y(y),   \, y=1,2,\ldots\label{eq:OurVersionOfTweediesFormula}
\end{align}
Moreover, 
\begin{align}
\rho^{\mathsf{Po}}_Y(y)&=  {\bf v} \boldsymbol{\cdot} \left[ \rho^{\mathsf{Po}, \, \mathsf{scale}}_Y(y), \rho^{\mathsf{Po}, \mathsf{d.c}}_Y(y)\right] \label{eq:ScoreComponents}     \\
&=  {\bf v} \boldsymbol{\cdot} \left[ \rho^{\mathsf{Po}, \, \mathsf{scale}}_Y(y), - \rho^{\mathsf{Po}, \mathsf{bw.d} }_Y(y) \right]   \label{eq:ScoreFUnctionMixWithBackward}   \\
&=-\rho^{\mathsf{Po}, \mathsf{fw.d} }_Y(y).  \label{eq:ScoreViaForwarDifference}
\end{align} 
\end{theorem} 
\begin{IEEEproof}
We start by re-writing the conditional expectation using the TGR formula \eqref{eq:COnditionalExpectation}
\begin{align}
\E \left[aX+\lambda| Y=y \right]%&=  \frac{ (y+1) P_Y(y+1)}{P_Y(y)}\\
%&=y- y+\frac{ (y+1) P_Y(y+1)}{P_Y(y)}\\
&= y+\frac{ (y+1) P_Y(y+1)- y P_Y(y)}{P_Y(y)},  \label{eq:ScoreFUnctionREP}
\end{align}
and work with the second term in \eqref{eq:ScoreFUnctionREP}.
Next, to show \eqref{eq:ScoreViaForwarDifference} observe that by using the definition of the forward difference we have that 
\begin{align}
\frac{ (y+1) P_Y(y+1)- y P_Y(y)}{P_Y(y)}=
\rho^{\mathsf{Po}, \mathsf{fw.d} }_Y(y).
\end{align}
To show  \eqref{eq:OurVersionOfTweediesFormula} and \eqref{eq:ScoreComponents}  use the identity in  \eqref{eq:GradientTypeConnection} 
\begin{align}
\frac{ (y+1) P_Y(y+1)- y P_Y(y)}{P_Y(y)}&= - \frac{ a \frac{ \rm d}{ { \rm d} a } P_Y(y) +\lambda \frac{\rm d }{{\rm d} \lambda  }  P_Y(y)  }{P_Y(y)} \\
&=  - {\bf v} \boldsymbol{\cdot}  \left[ \rho^{\mathsf{Po}, \, \mathsf{scale}}_Y(y), \rho^{\mathsf{Po}, \mathsf{d.c}}_Y(y)\right]\\
&=   - {\bf v} \boldsymbol{\cdot} \nabla_{\bf v} \log  P_Y(y)\\
&=\rho^{\mathsf{Po}}_Y(y). 
\end{align}  

Finally, to show \eqref{eq:ScoreFUnctionMixWithBackward}  use the identity in \eqref{eq:DarkCurrentDerivative}  to see that
\begin{align}
 \rho^{\mathsf{Po}, \mathsf{d.c}}_Y(y)&= \frac{    P_Y(y-1)- P_Y(y)   }{P_Y(y)}= -  \frac{    \mathsf{D}^{\mathsf{bw}}_1\left[ P_Y(y) \right]   }{P_Y(y)}.
 \end{align} 
This concludes the proof. 
\end{IEEEproof}

Unlike for continuous random variables, there are several definitions of a score function for discrete random variables. Specifically,  the following definitions of score functions for a  random variable $W$ supported on non-negative integers  have been proposed in \cite{kagan2001discrete}, \cite{kontoyiannis2005entropy}  and \cite{johnson2017bruijn}, respectively:  for  $w=0,1,2, \ldots $
\begin{align}
\rho^{\mathsf{K}}_W(w)&= \frac{P_W(w-1)-P_W(w)} {P_W(w)}   ,\label{eq:Score-K}\\
\rho_{W}^{\mathsf{KHJ}}(w)&=   \frac{ (w+1) P_W(w+1)}{ \E[W] P_W(w)}- 1,  \,   \label{eq:Score-KHJ}\\ 
\rho_{W}^{\mathsf{JG}}(w)&= w \E[W] \frac{P_W(w-1)-P_W(w)} {P_W(w)}-1.  \,     \label{eq:Score-JG}
\end{align} 

%We conclude this section by relating these definitions of score functions to the once proposed in  Theorem~\ref{thm:Tweedie'sFormulaPoisson}.
By letting $U=aX+\lambda$ and $\E[U]=a \E[X]+\lambda$ we observe the following relationship between  score functions proposed in Theorem~\ref{thm:Tweedie'sFormulaPoisson} and score functions in \eqref{eq:Score-K}, \eqref{eq:Score-KHJ} and  \eqref{eq:Score-JG}:  
\begin{align}
\rho^{\mathsf{Po}, \mathsf{d.c}}_Y(y)&=\rho^{\mathsf{K}}_Y(y)=\frac{\rho_{Y}^{\mathsf{JG}}(y) +1 }{ y \E[U] } ,\\
 \rho^{\mathsf{Po}}_Y(y) &= \E[U]  \rho_{Y}^{\mathsf{KHJ}}(y)+\E[U] -y.
\end{align}

The score function has an intimate relationship with \emph{the Fisher information}.  Motivated by our gradient definition of the score function in \eqref{eq:GradientDefintionoPoisson} we  define the following version of the Fisher information of $Y$ for the Poisson noise case:
\begin{align}
J^{\mathsf{Po}}(Y) = \E \left[ \left( \rho^{\mathsf{Po}}_Y(Y) \right)^2 \right]. 
\end{align} 
We can now show the following result between the Fisher information and the minimum mean squared error (MMSE) where the latter is defined as
\begin{align}
\mmse(X|Y)=\mathbb{E} \left[ (X-\E[X|Y])^2 \right]. 
\end{align} 
\begin{theorem}\label{thm:BrownsIdentityPoisson}  Let $Y=\mathcal{P}(aX+\lambda)$. Then, for every $a>0,\lambda \ge 0$
\begin{align}
\mmse(X|Y)= \frac{ a\E \left[  X  \right] +\lambda  - J^{\mathsf{Po}}(Y)}{a^2}  .  \label{eq:PoissonBrownIdentity}
\end{align}
\end{theorem} 
\begin{IEEEproof}
We start with the definition of the MMSE
\begin{align}
&\mmse(X|Y) \notag\\
&= \mathbb{E} \left[ (X-\E[X|Y])^2 \right]\\
& = \E \left[  \left( X- \frac{ Y+\rho^{ \mathsf{Po}}_Y(Y) -\lambda}{a}  \right )^2 \right] \label{eq:InsertScoreFunction}\\
&= \E \left[  \left( X- \frac{ Y-\lambda}{a}  \right )^2 \right]-2    \E \left[  \left( X- \frac{ Y -\lambda}{a}  \right ) \frac{\rho^{ \mathsf{Po}}_Y(Y)}{a} \right]\notag\\
&+\frac{1}{a^2} J^{\mathsf{Po}}(Y)\\
&= \E \left[  \left( X- \frac{ Y-\lambda}{a}  \right )^2 \right] \notag\\
&-2    \E \left[  \left( \E[X|Y]- \frac{ Y -\lambda}{a}  \right ) \frac{\rho^{ \mathsf{Po}}_Y(Y)}{a} \right]+\frac{1}{a^2} J^{\mathsf{Po}}(Y) \label{eq:LawOfTotalExpectation}\\
&= \E \left[  \left( X- \frac{ Y-\lambda}{a}  \right )^2 \right]-2  \frac{1}{a^2} J^{\mathsf{Po}}(Y)  +\frac{1}{a^2} J^{\mathsf{Po}}(Y) \label{eq:PlugInagain} \\
&= \frac{ a\E \left[  X  \right] +\lambda}{a^2} - \frac{1}{a^2} J^{\mathsf{Po}}(Y),
\end{align} 
where \eqref{eq:InsertScoreFunction} follows by using \eqref{eq:OurVersionOfTweediesFormula}; \eqref{eq:LawOfTotalExpectation} follows from the law of total expectation;   \eqref{eq:PlugInagain} follows by using the identity in \eqref{eq:OurVersionOfTweediesFormula}; and \eqref{eq:PlugInagain} follows by using 
\begin{align}
\E \left[  \left( X- \frac{ Y-\lambda}{a}  \right )^2 \right] &=  \E \left[   \mathbb{V} \left( \frac{ Y-\lambda}{a}  \mid X \right) \right] \notag\\
&=  \frac{1}{a^2} \E \left[   \mathbb{V} \left( Y\mid X \right) \right]= \frac{1}{a^2} \E \left[  aX +\lambda \right]. 
\end{align} 
This concludes the proof.  
\end{IEEEproof} 

\begin{remark} \label{rem:BrownGaussian}A Gaussian analog  of the identity in \eqref{eq:PoissonBrownIdentity} has been shown by  Brown in \cite{brown1971} 
\begin{align}
\mmse(V|Y_\mathsf{G})=  \frac{ \sigma^2-  \sigma^4 J(Y_\mathsf{G})}{a^2},
\end{align} 
where the Fisher information is given by 
\begin{align}
J(Y_\mathsf{G})=\E \left[ \left(  \frac{\rm d}{ {\rm d} y} \log \left(f_{Y_\mathsf{G}}(Y_\mathsf{G}) \right)  \right)^2\right].
\end{align}
\end{remark} 

As an example, by using that $J^{\mathsf{Po}}(Y) \ge 0$,  the expression in \eqref{eq:PoissonBrownIdentity}  leads to the bound
\begin{align}
\mmse(X|Y) \le \frac{a \E[X] +\lambda}{a^2}. \label{eq:BoundLargea}
\end{align} 
For the case of $\lambda=0$ and under some mild conditions on the pdf of $X$, the bound in \eqref{eq:BoundLargea} was shown to be order tight in \cite[Theorem~4]{dytso2020class}.   For upper and lower bounds and some other results on the MMSE of Poisson noise  the interested reader is referred to \cite{dytso2020class}.

 \subsection{Analytical Properties of the Conditional Expectation}
 \label{sec:AnaltyticalPropertiesOfConditionalExpecatation}
 
 In this section, we study how the conditional expectation behaves as a function of $y$ and  as  a function of the channel parameters $(a,\lambda)$.  

\begin{theorem}\label{thm:MonotonicityConditionalExpecation} Let $Y=\mathcal{P}(a X+\lambda)$. Then, for every fixed $a>0, \lambda \ge 0$,   the mapping $y  \to \E[X|Y=y]$ is non-decreasing. 
\end{theorem} 
 
  \begin{IEEEproof}
   To show that the expected value is non-decreasing let $U=aX+\lambda$ and consider
\begin{align}
P_Y(k) &= \frac{1}{k!} \E \left[ U^k \eu^{-U} \right]\\
&\le  \frac{1}{k!} \sqrt{ \E \left[ U^{k+1} \eu^{-U} \right]  \E \left[ U^{k-1} \eu^{-U} \right] } \label{eq:CauchySchwartz}\\
&=  \frac{1}{k!} \sqrt{ (k+1)! P_Y(k+1)   (k-1)! P_Y(k-1)}\\
&=   \sqrt{ \frac{k+1}{k} P_Y(k+1) P_Y(k-1)},  \label{eq:BoundOnPMFOutput}
\end{align}
where  in \eqref{eq:CauchySchwartz} we have used the Cauchy-Schwarz inequality.  Now applying the bound in  \eqref{eq:BoundOnPMFOutput}  to the TGR formula in \eqref{eq:COnditionalExpectation} we have that 
\begin{align}
\E[U|Y=y]&= \frac{(y+1) P_Y(y+1)}{P_Y(y)} \\
&\ge    \frac{(y+1)  \frac{y}{y+1} P_Y^2(y)}{P_Y(y) P_Y(y-1)}\\
&=    \frac{y P_Y(y)}{ P_Y(y-1)}\\
&= \E[U|Y=y-1].
\end{align} 
This shows that the conditional expectation is a non-decreasing function of $y$. 
\end{IEEEproof} 

 We note that the conditional expectation of  a non-degenerate random variable $X$ may not be strictly increasing and can increase only once.  Consider an example of a Bernoulli random variable and $\lambda=0$  for which
\begin{align}
\E[X|Y=y]=  \left \{\begin{array}{ll} \frac{  p \eu^{-a}}{ q+p \eu^{-a}  }  & y=0  \\ 
1 & y > 0 
\end{array}\right. . 
\end{align} 
For the Gaussian noise setting this, however, is not the case and the conditional expectation is a strictly increasing function for non-degenerate random variables (see Remark~\ref{rem:HiherOrderDerivatives} below). 

From Fig.~\ref{fig:PoissonConditionalExpectation} and Fig.~\ref{fig:ConditionalExpectationExponential} we observe that the conditional expectation with a lower dark current dominates the conditional expectation with a larger dark current. 
 This observation  holds in general  and is formally shown next. 

\begin{theorem}\label{thm:MonotonictyOfConditionalExp}  For  $Y=\mathcal{P}(a X +\lambda)$ and $U=aX+\lambda$ let ${\bf v}=[a,\lambda]$. 
 Then, for every $a>0, \lambda>0$ and  $y=0,1, \ldots$
\begin{align}
&{\bf v} \boldsymbol{\cdot}  \nabla_{{\bf v}} \E \left[X|Y=y \right] \notag\\
 & \qquad=  a \frac{ \rm d}{{\rm d} a } \E \left[X|Y=y \right] +   \lambda \frac{ \rm d}{{\rm d} \lambda } \E \left[X|Y=y \right] \notag \\
 & \qquad= - a  \mathbb{V}(X|Y=y), \label{eq:GradientOfConditionalExpectation}
%  
%  \left \{  \begin{array}{ll} 0 &  y=0\\
%  -  y  a \mathbb{V}(X|Y_{\lambda}=y-1)  & y \ge 1   \end{array} \right. ,  
%  
\end{align}
where 
\begin{align}
a \frac{ \rm d}{{\rm d} \lambda } \E[  X|Y=y] =    \left \{  \hspace{-0.2cm} \begin{array}{ll}    0 & y=0 \\
-  y      \frac{ \mathbb{V}(U|Y=y-1)}{  (\E \left[U|Y=y-1 \right])^2 }   & y \ge 1
\end{array}  \right.  .  \label{eq:DerivativeOfDarkCurrentConditionalExpectation}
\end{align} 
 More generally, for every positive integer $k$  
\begin{align}
&{\bf v} \boldsymbol{\cdot} \nabla_{{\bf v}} \E \left[U^k |Y=y \right]  \notag\\
& \qquad =a \frac{ \rm d}{{\rm d} a } \E[  U^k|Y=y] + \lambda  \frac{ \rm d}{{\rm d} \lambda } \E[  U^k|Y=y] \\
& \qquad = k  \E[  U^k|Y=y]- \E[  U^{k+1}|Y=y]    \notag\\
&\qquad \quad  + \E[  U^k|Y=y] \E[  U|Y=y],  \label{eq:GradientOfCOnditionalExpectation}
\end{align}
where 
\begin{align}
& \frac{ \rm d}{{\rm d} \lambda } \E[  U^k|Y=y]  \notag\\
&\qquad =  \left \{  \hspace{-0.2cm} \begin{array}{ll} k \E[ U^{k-1} |Y=0]  &  y=0\\   \frac{ (y+k) \E[ U^{k-1} |Y =y]  \E[ U |Y =y-1]   - y \E[ U^{k} |Y =y]   }{ \E[ U |Y =y-1] } & y \ge 1   \end{array} \right.  . \label{eq:DerivativeOfHiherMoments}
\end{align}

\end{theorem} 

\begin{IEEEproof}
See Appendix~\ref{app:thm:MonotonictyOfConditionalExp}. 
\end{IEEEproof} 

As a consequence of Theorem~\ref{thm:MonotonictyOfConditionalExp} we have the following corollary. 
\begin{corollary}\label{cor:LimitAndMOnotonicity} Let $Y_{\lambda}=\mathcal{P}(a X +\lambda)$ where $X$ is a non-degenerate random variable.  Then, for a fixed $y \ge 1$, the mapping $\lambda  \to \E \left[X|Y_{\lambda}=y \right]$ is strictly decreasing.   Moreover,  
\begin{align} 
\lim_{\lambda \to \infty}  \E \left[X|Y_{\lambda}=y \right]= \E[X|Y_{\kappa}=0], \, y=0,1,\ldots \label{eq:LambdaGoesToINfinity}
\end{align} 
where  $\kappa \ge 0$ can be  arbitrary.  
\end{corollary}
\begin{IEEEproof}
The fact that  $\lambda \to \E \left[X|Y_{\lambda}=y \right]$ is strictly decreasing for $y \ge 1$ follows from \eqref{eq:DerivativeOfDarkCurrentConditionalExpectation}  by using the fact that the variance of a non-degenerate random variable is positive. 

The proof of \eqref{eq:LambdaGoesToINfinity} proceeds as follows: 
\begin{align}
\lim_{\lambda \to \infty} \E[X|Y_\lambda=y] &= \lim_{\lambda \to \infty} \frac{ \E \left[X  (a X+\lambda)^y \eu^{-(aX+\lambda)} \right]}{ \E \left[  (a X+\lambda)^y \eu^{-(aX+\lambda)} \right] }\\
&= \lim_{\lambda \to \infty} \frac{ \E \left[X   \left(  \frac{a X}{\lambda}+1 \right)^y \eu^{-aX} \right]}{  \E \left[  \left(  \frac{a X}{\lambda}+1 \right)^y \eu^{-aX} \right] } \label{eq:application of Dominated ConvergenceTheoremTOlimitDarkCurrent}\\
&=\frac{ \E \left[X   \eu^{-aX} \right]}{  \E \left[   \eu^{-aX} \right] } \\
&=  \E[X|Y_0=0]\\
&= \E[X|Y_\kappa=0], \label{eq:InsertingKappa}
\end{align}
where the exchange of the limit and expectation in \eqref{eq:application of Dominated ConvergenceTheoremTOlimitDarkCurrent} follows by using the dominated convergence theorem with  the following bound that holds for $\lambda >1$:
\begin{align}
X   \left(  \frac{a X}{\lambda}+1 \right)^y \eu^{-aX} \le  (y+1)^{y+1} \eu^{-y};
\end{align} 
the proof  of the above bound follows along the same lines as that of \eqref{eq:TailOverAllSitributions}; and $\kappa \ge 0$ in \eqref{eq:InsertingKappa}  can be taken arbitrary since the conditional expectation at $y=0$  does not depend on the dark current parameter.   This concludes the proof. 
\end{IEEEproof}

%An interesting consequence of  combining the derivative expression in \eqref{eq:DerivativeOfDarkCurrentConditionalExpectation} and the limit in \eqref{eq:LambdaGoesToINfinity} is the following integral representation of the conditional expectation:  let $Y_{\bf v}=\mathcal{P}(aX+\lambda)$ where ${\bf v}=[a,\lambda]$, then 
%\begin{align}
%\E[X| Y_{{\bf v}_0}=y]=\E[X|Y_{{\bf v}_\infty}=0] +    a  \oint_{\mathcal{C}}  \hspace{-0.1cm} \mathbb{V}(X|Y_{\bf v}=y) \cdot {\rm d} {\bf v},
%\end{align}
%where $\mathcal{C}$ is a path between ${\bf v}_1=[a_0,\lambda_0]$ and $  {\bf v}_\infty=[ a_1, \infty]$. 

\begin{remark}\label{rem:HiherOrderDerivatives}  For the Gaussian noise channel in \eqref{eq:GaussianChannel} the analog of the   identity in \eqref{eq:GradientOfConditionalExpectation}  is the following formula shown in  \cite{hatsell1971some}:
\begin{align}
  \sigma^2 \frac{\rm d}{ {\rm d}y} \E[V|Y_\mathsf{G}=y]=  \mathbb{V}(V|Y_\mathsf{G}=y).
 \end{align}
 Moreover, the  Gaussian analog of the identity in \eqref{eq:GradientOfCOnditionalExpectation} is 
 \begin{align}
& \sigma^2  \frac{\rm d}{ {\rm d}y} \E[V^{n-1}|Y_\mathsf{G}=y] \notag\\
 &= \E[V^n|Y_\mathsf{G}=y]-\E[V^{n-1}|Y_\mathsf{G}=y] \E[V|Y_\mathsf{G}=y], 
 \end{align} 
 shown in \cite{jaffer1972note}.  
 These two observations further support the view that  differentiation with respect to $a$ and $\lambda$ in  a Poisson channel is the analog of differentiation with respect to $y$ for the Gaussian channel. 
\end{remark}

 We conclude this section by  using the monotonicity of conditional expectation to show  an inequality that has the flavor of the reverse Jensen's inequality.
\begin{lem}\label{lem:reverseJensen} Let   $Y=\mathcal{P}(X)$. Then, for every positive integer $k$ and  non-negative integer $y$
\begin{align}
 \E[X|Y=y]  &\le \E^{\frac{1}{k}} \left[X^k|Y=y \right]  \le \E \left[X|Y=y+k-1 \right].  \label{eq:referseJenseInequality}
\end{align}
\end{lem} 
\begin{IEEEproof}
Using the monotonicity of the conditional expectation observe the following inequalities: 
\begin{align}
  \left( \E \left[X|Y=y \right]  \right)^k  &\le   \prod_{i=0}^{k-1} \E[X|Y=y+i]   \notag\\
  &\le  \left( \E[X|Y=y+k-1] \right)^k.
\end{align}
The proof is now concluded by using the identity in  \eqref{eq:productIdentityForConditionalMoments} .
\end{IEEEproof}

Observe that the first inequality in \eqref{eq:referseJenseInequality} could have been shown by using Jensen's inequality.

\subsection{Uniqueness of the Conditional Expectation with Respect to the Input Distribution} 
\label{sec:UniquenessOfConditionalExpecatation}
In this section, we are interested in  whether the knowledge of  $\E[U|Y=y]$ for all $y$  uniquely determines the input distribution $P_U$.   

We begin by showing an auxiliary result about the conditional distribution, which is of  independent interest,  that will be used to show the uniqueness of the conditional expectation with respect to the input distribution.

\begin{theorem}\label{thm:DeterminanceOfthConditional} Fix some non-negative integer  $y$  and let $U_y$ be distributed according to $P_{U|Y=y}$ where $Y=\mathcal{P}(U)$.   Denote by $ \{m_k \}_{k=1}^\infty$ the sequence of integer moments of $U_y$ (i.e.,  $m_k= \E[U_y^k]$).  

Then,  the distribution of $U_y$ is uniquely determined by the sequence  $\{m_k \}_{k=1}^\infty$. 
\end{theorem} 
\begin{IEEEproof}
Clearly, we have that  $m_k<\infty$ for all $k>1$, which follows from the inequality in Lemma~\ref{lem:reverseJensen}. In the rest of the proof,   we seek to determine whether the moments of $U_y$ are unique.   Since $U_y \ge 0$,  this is a classical  Stieltjes moment problem \cite{shohat1943problem}. The following sufficient condition for the uniqueness of moments was given by  Carleman \cite{stoyanov2013counterexamples}:  the moments of $U_y$ are unique if 
\begin{align}
\sum_{k=1}^\infty  \left(\E[ U_y^k ] \right)^{- \frac{1}{2k}} =\infty.  \label{eq:CarlemanCondtion}
\end{align}

Next, using the upper bound in \eqref{eq:TailOverAllSitributions} observe the following inequality:  
\begin{align}
\E[ U_y^k ] & = \frac{(y+k)!}{ y!}  \frac{P_Y(y+k)}{P_Y(y)}. \\
& \le   \frac{(y+k)!}{ y!}  \frac{   \frac{1}{(y+k)!}  (y+k)^{y+k} \eu^{-(y+k)}  }{P_Y(y)}\\
%& =   \frac{   (y+k)^{y+k} \eu^{-(y+k)}  }{  y! P_Y(y)}\\
&= c_y     (y+k)^{y+k} \eu^{-k}, \label{eq:boundOnMoments} 
\end{align}
where in the last step we have defined
\begin{align}
c_y=  \frac{    \eu^{-y}  }{  y! P_Y(y)}. 
\end{align}

Now, applying the bound in \eqref{eq:boundOnMoments} to the summation in \eqref{eq:CarlemanCondtion}
\begin{align}
\sum_{k=1}^\infty  \left(\E[ U_y^k ] \right)^{- \frac{1}{2k}}  \ge \eu^{ \frac{1}{2}}     \sum_{k=1}^\infty   \frac{1}{c_y^{\frac{1}{2k}}  (y+k)^{\frac{1}{2k}}} ,
\end{align}
since $y$ and $c_y$ are fixed the above sum diverges by the comparison test. Therefore, the Carleman condition in \eqref{eq:CarlemanCondtion} is satisfied and the moments determine the distribution. This concludes the proof.  
\end{IEEEproof}

\begin{remark} To the best of our knowledge, no  Gaussian analog of Theorem~\ref{thm:DeterminanceOfthConditional}  has ever been presented. Nonetheless,  for the Gaussian noise case, a  version of Theorem~\ref{thm:DeterminanceOfthConditional}  can be shown by using  \cite[Proposition~6]{GuoMMSEprop}, where it was shown that  $P_{V|Y_{\mathsf{G} }=y}$ is  sub-Gaussian for every fixed $y$ regardless of the  distribution of the input $V$, together with  Carleman's conditions for real-valued random variables.     
We note, however, that in  general for the Poisson case the conditional distribution $P_{U|Y}$ does not enjoy the sub-Gaussianity property. To see this,    recall that  sub-Gaussian random variables have moments that grow at a rate of at most $C^k  \sqrt{k!}$ for some fixed constant $C>0$ \cite{buldygin1980sub}. Now,  consider $U \sim \mathsf{Gam}(1, 1)$ in which case using Table~\ref{table:CEexamples} we have that
\begin{align}
\E[U|Y=y]=\frac{y+1}{2}, y=0,1,\ldots
\end{align} 
 and using \eqref{eq:productIdentityForConditionalMoments}  the higher order moments  of $U|Y=y$ are given by
\begin{align}
m_k= \frac{(y+k)!}{2^k y!}. 
\end{align}
This, however, for every fixed $y$, grows faster  than $C^k  \sqrt{k!}$ and so $P_{U|Y=y}$ is not sub-Gaussian. 
\end{remark}

The next result establishes the uniqueness of the conditional expectation.

\begin{theorem}\label{thm:UniquenssOfConditionalExpectation}  Let  $Y_1 = \mathcal{P} (U_1)$ and   $Y_2 = \mathcal{P} (U_2)$.   Then,  the conditional expectation is a bijective operator of the input distribution; that is
\begin{align} 
\E[U_1 | Y_1=y]= \E[U_2 | Y_2=y], \,\forall y  \,    \Longleftrightarrow  \,   P_{U_1}= P_{U_2}. 
\end{align} 
\end{theorem} 
\begin{IEEEproof}
Suppose that  $P_{U_1}= P_{U_2}$; then it is immediate that  
\begin{align}
 \E[U_1 | Y_1=y]= \E[U_2 | Y_2=y],  \,\forall y  .
\end{align} 
 Therefore, it remains to show the other direction. 

Next suppose that $\E[U_1 | Y_1=y]= \E[U_2 | Y_2=y], \forall y  $.    Then,  using the identity in \eqref{eq:productIdentityForConditionalMoments} we have that for all   integers $k \ge 1$
\begin{align}
\E[U_1^k | Y_1=y]= \E[U_2^k | Y_2=y], \,  \forall y.  \label{eq:AllHigherConditonalMoments}
\end{align} 
Using Theorem~\ref{thm:DeterminanceOfthConditional}, the expression in \eqref{eq:AllHigherConditonalMoments} implies that   
\begin{align} 
P_{U_1|Y_1=y}= P_{U_2|Y_2=y},  \forall y.   \label{eq:COnditionalsAreEqualForAlly}
\end{align}
Now the equality in \eqref{eq:COnditionalsAreEqualForAlly} implies that $P_{U_1}= P_{U_2}$. To see this choose some measurable set $A \subset \mathbb{R}$ and observe that
\begin{align}
P_{U_1}(A)&= \E[  1_{A}(U_1) ]\\
&= \E \left[  \E \left[ 1_{A}(U_1) | Y_1 \right]  \right ]\\
&= \E \left[  \E \left[ 1_{A}(U_2) | Y_2 \right]  \right ]\\
&=P_{U_2}(A). 
\end{align} 
This concludes the proof.   %{\color{red} Give more details on the last step here} 
\end{IEEEproof}

\begin{remark}\label{rem:GaussianUniqueness}
For the Gaussian noise channel, the uniqueness of the conditional expectation has been established in \cite[Appendix~B]{FunctionalPropMMSE}.   We note, however,  that our proof and the proof in \cite{FunctionalPropMMSE} are very different.  Moreover, to the best of our knowledge, there is no single unifying approach for demonstrating uniqueness results for the conditional expectation with respect to the input distribution. 
\end{remark}

There are several interesting consequences of Theorem~\ref{thm:UniquenssOfConditionalExpectation}:
\begin{itemize}[leftmargin=*]
\item   (Strict Concavity of the MMSE).  Consider $Y=\mathcal{P}(aX+\lambda)$ for some $X$  and the corresponding MMSE of estimating $X$ from $Y$
\begin{align}
\mmse(X|Y)=\E \left[ (X-\E[X|Y])^2 \right].  
\end{align} 
The uniqueness of the conditional expectation implies  that $P_X \to \mmse(X|Y)$ is a strictly concave mapping.  This follows by applying  \cite[Theorem~1]{FunctionalPropMMSE}  where it was established  that the MMSE is strictly concave provided that the conditional expectation uniquely determines the input distribution; 

\item  (Least Favorable Distributions).      A distribution $P_X$ is said to be \emph{least favorable} with respect to the MMSE and some parameter space $\Omega$ if
\begin{align} 
P_X \in \arg \max_{P_X: X \in \Omega} \mmse(X|Y).  \label{eq:lookingFOrLeastFavorableDistributions}
\end{align}
The uniqueness of the conditional expectation, which implies that the objective function in \eqref{eq:lookingFOrLeastFavorableDistributions} is strictly convex,  guarantees that the least favorable prior distribution is unique. Moreover, this also implies  uniqueness of the minimax  estimator of a deterministic parameter $ \theta \in\Omega$ of the risk 
\begin{align}
R(\theta, \hat{\theta})=\E \left[   \left(\theta -\hat{\theta}(Y) \right)^2 \right], \text{ where $Y=\mathcal{P}(a \theta +\lambda)$.} 
\end{align}
 The interested reader is  also referred to \cite{dytso2018structure} where conditions for the least favorable prior to be  binary  have been shown; 
 
\item (Empirical  Bayes).  Consider an  independent and identically sequence (i.i.d.) of input random variables $ \{ X_i \}$ and a corresponding  output sequence $ \{ Y_i \}$ where $Y_i =\mathcal{P}(X_i)$.   Now consider the expression for the TGR formula  given by 
\begin{align}
\E[X|Y=y]= \frac{(y+1) P_Y(y+1)}{ P_Y(y)}. 
\end{align} 
Because the conditional estimator in the TGR formula depends only on the marginal of the output $Y$, from the  $Y_i$ observations we can build an empirical distribution $\widehat{P}_Y$ and construct an empirical version of the conditional expectation
\begin{align}
\widehat{\E}[X|Y=y]= \frac{ (y+1) \widehat{P}_Y(y+1)}{\widehat{P}_Y(y)}.
\end{align} 
In other words, we are able to approximate the optimal estimator without the knowledge of the prior distribution on $X$. This remarkable procedure was first developed by Robbins in \cite{robbins1956empirical}.  The uniqueness result of Theorem~\ref{thm:UniquenssOfConditionalExpectation}  implies that the empirical Bayes procedure is not  only producing an estimate of the conditional expectation but  is also (for free!) producing an estimate of the distribution of $X$.  Therefore, it will be interesting to characterize the exact inverse transform relationship between $P_X$ and $\E[X|Y=y]$; and
\item  (Uniqueness of the Ratio of Derivatives of the Laplace Transform).  Another interesting byproduct of Theorem~\ref{thm:UniquenssOfConditionalExpectation}  is that the sequence of ratios of derivatives of  the Laplace transform evaluated at one, that is 
\begin{align}
\frac{  \mathcal{L}_{U}^{(k+1)}(1) }{ \mathcal{L}_{U}^{(k)}(1) }, k=0, 1, \ldots,  \label{eq:RatioLaplace}
\end{align} 
completely determines the distribution $P_U$.  To see that sequence in  \eqref{eq:RatioLaplace}  completely determines $P_U$,  use the Laplace identity in \eqref{eq:LaplaceRepresentation}.
\end{itemize}

\subsection{Bounds on the Conditional Expectation} 
\label{sec:BoundsOnTheConditionalExpectation}
In this section, we are interested in determining if the rate of growth of the conditional expectation as a function of $y$.
Specifically, we are concerned with the question of whether the conditional expectation can have a super-linear growth. 

\begin{remark}\label{rem:BoundOnConditionalExpectation}  Upper bounds on the conditional expectation  for the Gaussian noise channel have been previously reported in  \cite[Lemma 4]{FunctionalPropMMSE} and \cite[Proposition 1.2]{fozunbal2010regret} where it was demonstrated that   $ | \E \left[V | Y_{\mathsf{G}}=y \right]|=O(|y|).$\footnote{For two non-negative functions $f$ and $g$ over $\mathbb{R}$, we say that $f(x)=O(g(x))$ if and only if $\limsup_{x \to \infty} \frac{f(x)}{g(x)}  <\infty$.} That is, in the Gaussian noise case the conditional expectation can have at most linear growth as a function $y$. 
\end{remark} 

The question of whether the conditional expectation can grow faster than a linear function can be reduced to  answering whether the following supremum is finite: 
\begin{align}
\sup_{y \ge 0} \frac{\E[ U| Y=y]}{ y+1} =  \sup_{y \ge 0} \frac{P_Y(y+1)}{P_Y(y)},  \label{eq:ratioOfPMFS}
\end{align} 
where we have used the TGR formula.

\begin{remark}
We note that it is possible to construct an example of a discrete random variable $W$ supported on non-negative integers such that 
\begin{align}
 \sup_{n \ge 0} \frac{P_W(n+1)}{P_W(n)}= \infty.
\end{align} 
Take an arbitrary discrete distribution $Q$ supported on non-negative integers and define
\begin{align}
P_W(n) &=c \frac{Q(n)}{n}  \text{ for $n$ even;  and} \\
 P_W(n)&=c Q(n-1) \text{ for $n$ odd}  ,
\end{align} 
where $c$ is the normalization constant. Then,
\begin{align}
 \frac{P_W(n+1)}{P_W(n)}=n,
\end{align}  
and $ \sup_{n \ge 0} \frac{P_W(n+1)}{P_W(n)}=\infty$. 
This construction suggests that we cannot rule out the possibility of \eqref{eq:ratioOfPMFS} being infinite.
\end{remark} 

  The next result provides a sufficient conditions for the finiteness of the supremum \eqref{eq:ratioOfPMFS} in terms of properties of the Laplace transform of the input random variable.

\begin{theorem}\label{thm:AssimptoticUppoerBoun} Let $Y=\mathcal{P}(U)$.   Then,  the following  hold:
\begin{enumerate}
\item (Lower Envelope).   For every $U$
\begin{align}
\liminf_{y \to \infty}  \frac{\E[ U| Y=y]}{ y+1}  <1.  \label{eq:LowerEvelop}
\end{align} 
\item (Upper Evelope). 
\begin{align}
\sup_{y \ge 0} \frac{\E[ U| Y=y]}{ y+1} <\infty,  \label{eq:LinearBigOboundA}
\end{align}
if and only if
\begin{align}
\limsup_{n \to \infty} \left|  \frac{ \mathcal{L}_{U}^{(n+1)}(1)  }{ (n+1) \mathcal{L}_{U}^{(n)}(1)  } \right|<\infty. \label{eq:LimitSupermumCondition}
\end{align} 
\item  Let $U$ be such  that the following limit exists: 
\begin{align}
\lim_{n \to \infty} \left|  \frac{ \mathcal{L}_{U}^{(n+1)}(1)  }{ (n+1) \mathcal{L}_{U}^{(n)}(1)  } \right|= L.  \label{eq:AssumptionAboutTheLimit}
\end{align} 
Then,   
\begin{align}
L &<\infty, \text{ and } \\ 
\sup_{y \ge 0} \frac{\E[ U| Y=y]}{ y+1} &<\infty.    \label{eq:LinearBigObound}
\end{align}
Moreover, 
\begin{align}
 \lim_{n \to \infty} \frac{\E[ U| Y=y]}{ y+1}=  L= \lim_{y \to \infty}   \left| \frac{  \mathcal{L}_{U}^{(y)}(1)}{y!} \right|^{ \frac{1}{y}} \le 1. 
\end{align} 
\end{enumerate}

\end{theorem}
\begin{IEEEproof}
To show that \eqref{eq:LinearBigOboundA} is equivalent to \eqref{eq:LimitSupermumCondition}, note that 
\begin{align}
\sup_{y \ge 0} \frac{\E[ U| Y=y]}{ y+1} =  \sup_{y \ge 0}  \left|  \frac{ \mathcal{L}_{U}^{(y+1)}(1)}{ (y+1) \mathcal{L}_{U}^{(y)}(1)} \right|,
\end{align} 
and 
\begin{align}
 \sup_{y \ge 0}  \left|  \frac{ \mathcal{L}_{U}^{(y+1)}(1)}{ (y+1) \mathcal{L}_{U}^{(y)}(1)} \right|  \hspace{-0.05cm} < \infty \Leftrightarrow  \limsup_{n \to \infty} \left|  \frac{ \mathcal{L}_{U}^{(n+1)}(1)  }{ (n+1) \mathcal{L}_{U}^{(n)}(1)  } \right|  \hspace{-0.05cm}  <  \hspace{-0.05cm} \infty.
\end{align} 

To show the first and third part of the theorem  recall the following inequalities:
for a sequence of real numbers $ \{ a_n  \}_{n=1}^\infty$\begin{align}
 \liminf_{ n \to \infty }  \frac{|a_{n+1}|}{ |a_{n}| } & \le \liminf_{ n \to \infty } |a_n|^{  \frac{1}{n}} \notag\\
 &  \le \limsup_{ n \to \infty } |a_n|^{  \frac{1}{n}} \le  \limsup_{ n \to \infty }  \frac{|a_{n+1}|}{ |a_{n}| } .  \label{eq:LimSupSequence}
\end{align} 

Now let $|a_n|=  \left| \frac{1}{n!}  \mathcal{L}_{X}^{(n)}(1) \right|$. Therefore,  using the sequence of inequalities in \eqref{eq:LimSupSequence} we have that
\begin{align}
\liminf_{n \to \infty}  \frac{\E[ U| Y=n]}{ n+1}&=  \liminf_{n \to \infty} \left|  \frac{ \mathcal{L}_{U}^{(n+1)}(1)  }{ (n+1) \mathcal{L}_{U}^{(n)}(1)  } \right|  \\
&\le   \liminf_{n \to \infty}   \left| \frac{1}{n!}  \mathcal{L}_{U}^{(n)}(1) \right|^{ \frac{1}{n}} \label{eq:LiminfEquality} \\
& \le   \liminf_{n \to \infty}   \left( \frac{1}{n!}   n^n \eu^{-n} \right)^{ \frac{1}{n}} \label{eq:BoundOnLaplaceT} \\
&=1,
\end{align} 
where in \eqref{eq:BoundOnLaplaceT} we have used that  $|  \mathcal{L}_{U}^{(n)}(1)  | =  | \E[ U^n \eu^{-U} ] | \le    n^n \eu^{-n}$. 
Now, if the limit in \eqref{eq:AssumptionAboutTheLimit} exists, then the inequality in \eqref{eq:LiminfEquality} becomes equality which leads to the proof of 3). 

This concludes the proof. 
\end{IEEEproof}

\begin{remark}
An interesting future direction would be to investigate  whether  the condition in  \eqref{eq:AssumptionAboutTheLimit} 
is also necessary for finiteness or to find an input $U$ that provides a counterexample. 
\end{remark}

Regrettably the bound  \eqref{eq:LinearBigObound} is not explicit. The next theorem provides an explicit upper bound and shows that the rate of growth of a conditional expectation cannot exits $O( y \log y)$ for the inputs with finite mean. 

\begin{theorem}\label{Thm:BoundOnConditionalExpectation}  Let $Y=\mathcal{P}(U)$. Then,  for  $y=0,1,\ldots$ 
\begin{align}
\frac{\E[U|Y=y]}{2} &\le  (y+1) \log(y+1) - y  \E[\log(U)]   + \E[U] \notag\\
&+  \frac{1}{2 \eu}+1 . \label{eq:BoundEstiamtor}
\end{align}
\end{theorem}
\begin{IEEEproof}
See Appendix~\ref{app:Thm:BoundOnConditionalExpectation}. 
\end{IEEEproof}

\begin{figure}[h!]  
	\centering   
\begin{tikzpicture}

\begin{axis}[%
width=7.5cm,
height=5.7cm,
at={(1.011in,0.642in)},
scale only axis,
xmin=0,
xmax=10,
xlabel style={font=\color{white!15!black}},
xlabel={$y$},
ymin=0,
ymax=16.5,
ylabel style={font=\color{white!15!black}},
ylabel={$ \E[X|Y=y]$},
axis background/.style={fill=white},
xmajorgrids,
ymajorgrids,
legend style={legend cell align=left, align=left, draw=white!15!black,at={(0.9,0.9)}}
]

\addplot [color=black, mark=*, mark options={solid, black}]
  table[row sep=crcr]{%
0	1.07692307692308\\
1	1.84615384615385\\
2	2.61538461538461\\
3	3.38461538461538\\
4	4.15384615384615\\
5	4.92307692307692\\
6	5.69230769230769\\
7	6.46153846153846\\
8	7.23076923076923\\
9	8\\
10	8.76923076923077\\
};
%\addlegendentry{Bound in }

\addplot [color=black, mark=o, mark options={black}]
  table[row sep=crcr]{%
0	5.85060638725239\\
1	6.09431248863146\\
2	6.86126673377508\\
3	7.96801905250949\\
4	9.32744291045961\\
5	10.8882219039166\\
6	12.6164478721947\\
7	14.4880209025053\\
8	16.4849215053518\\
9	18.5931629795255\\
10	20.8015717906263\\
};
%\addlegendentry{data2}

\end{axis}

\end{tikzpicture}%
		\caption{Comparison of the bound in \eqref{eq:BoundEstiamtor} (clear circles)  to   $\E[U|Y=y]$ (filled circles) where $U \sim  \mathsf{Gam}(0.3,1.4 )$.   }
	\label{fig:BoundsOnCOnditionalEstiamtor}
\end{figure}
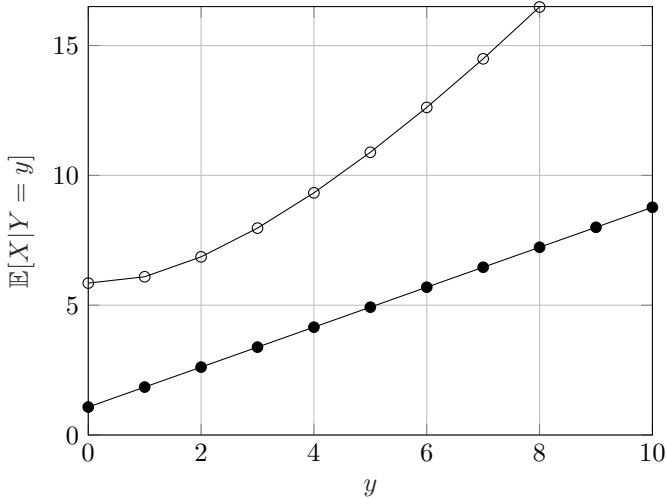

Fig.~\ref{fig:BoundsOnCOnditionalEstiamtor} compares a bound in \eqref{eq:BoundEstiamtor}  to    $\E[U|Y=y]$ where $U$ is according to a gamma distribution.

\section{Linear Estimation and  the Conditional Expectation}
\label{sec:LinearityProperties}
In this section, we study the properties of linear estimators. More specifically,  our interest lies in various questions of optimality of linear estimators such as:
\begin{enumerate} [leftmargin=*]
\item Under what input distributions are linear estimators optimal for squared error loss and Bregman divergence loss? Since the conditional expectation is an optimal estimator for the aforementioned  loss functions, this is equivalent to asking when is the conditional expectation a linear function of $y$; and 
\item If the linear estimators are approximately optimal,  can we say something about the input distribution?  In other words, we are looking for a quantitative refinement of 1). 
\end{enumerate}

\subsection{When is the Conditional Expectation Linear?}
\label{sec:LinearBoundAttained}

Linear estimators are important in estimation theory due to the simplicity of their implementation and analysis.   In the Gaussian noise case linear estimators are induced by the Gaussian input. In  Poisson noise the same role is played by the gamma distribution. 

   The next result provides sufficient and necessary conditions  on the  distribution  of $X$ and parameters of the Poisson transformation $(a,\lambda)$ that guarantee the linearity of $\E[X|Y]$. 

\begin{theorem}\label{thm:LinearityConditions} Let $Y=\mathcal{P}(a X+\lambda)$.  Then, 
\begin{align}
\E[X|Y=y]= b_1 y+b_2, \,  \forall y=0,1,\ldots
\end{align} 
 if and only if  the following two conditions hold:
 \begin{itemize}[leftmargin=*]
 \item $\lambda=0$;    and 
 \item 
$X \sim  \mathsf{Gam} \left(  \frac{1-ab_1}{b_1}, \frac{b_2}{ b_1} \right)$ for any $0 <b_1 <\frac{1}{a}$ and   $b_2>0$. 
\end{itemize}
\end{theorem}
\begin{IEEEproof} 
See Appendix~\ref{app:thm:LinearityConditions}.
\end{IEEEproof}

One of the ramifications of Theorem~\ref{thm:LinearityConditions} is that $\E[X|Y]$ is linear only if $\lambda=0$. In other words, in the presence of dark current, somewhat disappointingly,  linear estimators are not optimal for a large class of loss functions.   

The existence of an input distribution that results in linear estimators plays an important role in estimation theory.  For example, to provide performance guarantees minimum mean squared error is often upper bounded by analyzing a sub-optimal linear estimator.  
Hence, the existence of input distributions with linear conditional expectations  shows that such bounds  can be attained.

\begin{remark}  For $\lambda=0$,  Theorem~\ref{thm:LinearityConditions} could have been derived by using the fact that the gamma distribution is a  \emph{conjugate prior} distribution of the Poisson distribution \cite{diaconis1979conjugate, johnson1957uniqueness}.  This implies that for $Y=\mathcal{P}(aX)$  if $X \sim   \mathsf{Gam}(\alpha,\theta)$, then    $aX|Y=y \sim  \mathsf{Gam}\left( \frac{\alpha}{a}+1,\theta+y\right)$.   

Furthermore, the fact that the posterior distribution is gamma allows one to show that the conditional variance of $X$ given $Y$ is also linear and is given by 
\begin{align}
\mathbb{V}(X|Y=y)=     \frac{\theta +y}{ ( \alpha+a )^2 }. 
\end{align}
In contrast, for the Gaussian noise channel in \eqref{eq:GaussianChannel} the linear estimator is obtained if and only if  the  input is Gaussian (Gaussian is also a conjugate prior of Gaussian), but the conditional variance is constant and independent of the observation $y$. Specifically, for $V_\mathsf{G} \sim \mathcal{N}(0,1)$   related to $Y_\mathsf{G}$ through \eqref{eq:GaussianChannel} 
\begin{align}
\mathbb{V}(V_\mathsf{G}|Y_\mathsf{G}=y)= \frac{\sigma^2}{1+\sigma^2},  \forall y \in \mathbb{R}.
\end{align} 
\end{remark}  

%Another interesting property of the Gamma distribution is shown next.
%\begin{lem} Let $Y= \mathcal{P}(aX)$ and let $X\sim \mathsf{Gam}(\alpha,\theta)$. Then,
%\begin{align}
%\mathbb{V}(X|Y=y)=  \frac{1}{a^2}
%\end{align}
%\end{lem}
%\begin{IEEEproof}
%We first look at 
%\begin{align}
%\mathbb{V}(aX|Y=y)=   \E \left[ (aX)^2|Y=y \right]- \E \left[ aX|Y=y \right]
%\end{align}
%
%
%This expressions in \eqref{}, \eqref{} and expression in \eqref{}  we have that 
%\begin{align}
%&\E \left[ (aX)^2|Y=y \right]  \notag\\
%&=  \frac{a^2}{     \left(\alpha+ a\right)^{2}}   (\theta +y) (\theta+y+1)
%\end{align} 
%and
%\begin{align}
%&\E \left[ (aX)^2|Y=y \right]  \notag\\
% &=  (y+1)  \frac{P_Y(y+1)}{P_Y(y)}\\
%&=   (y+1)  \frac{   (-a)^{y+1}    \frac{ \alpha^\theta}{  \left(\alpha+ a\right)^{\theta+y+1} } { {-\theta} \choose y+1}  }{  (-a)^{y}    \frac{ \alpha^\theta}{  \left(\alpha+ a\right)^{\theta+y} } { {-\theta} \choose y}}\\
%&=   ((y+1)  \frac{  a  { {-\theta} \choose y+1}  }{     \left(\alpha+ a\right)  { {-\theta} \choose y}}\\
%&= - \frac{a}{     \left(\alpha+ a\right)}   (\theta+y)
%\end{align} 
%\end{IEEEproof}

To present an illustrative example of the effect of dark current on the conditional expectation we compute the conditional expectation  for $\lambda \ge 0$ of an input according to the exponential distribution, which is the gamma distribution with a unit shape parameter. 

\begin{lem}\label{lem:OutputExponential}  Let $Y=\mathcal{P}(aX+\lambda)$ and take  $X$ to be an exponential random variable of rate $\alpha$. Then, for every $a>0$ and $\lambda \ge 0$ 
\begin{subequations}
\begin{align}
&P_Y(0)=  \frac{\alpha \eu^{-\lambda}}{ \alpha+a},    \\
&P_Y(k)=\frac{\Gamma(k+1,\lambda)}{\Gamma(k+1)}-\frac{\Gamma(k,\lambda)}{\Gamma(k)} \notag\\
&+\frac{\eu^{\frac{\alpha}{a} \lambda}}{ \left( 1+\frac{\alpha}{a}  \right)^{k}} \left(   \frac{\Gamma \left( k, \lambda \left(\frac{\alpha}{a} +1 \right) \right)  }{\Gamma \left( k \right)   }  -     \frac{\Gamma \left( k+1, \lambda \left(\frac{\alpha}{a} +1 \right) \right)  }{\Gamma \left( k+1 \right)   \left( 1+\frac{\alpha}{a}  \right) }   \right),
\end{align}
\end{subequations}
\end{lem}
where  $\Gamma(\cdot, \cdot)$ is the upper incomplete gamma function. 
\begin{IEEEproof}
The proof follows via tedious but otherwise elementary integration. 
\end{IEEEproof}

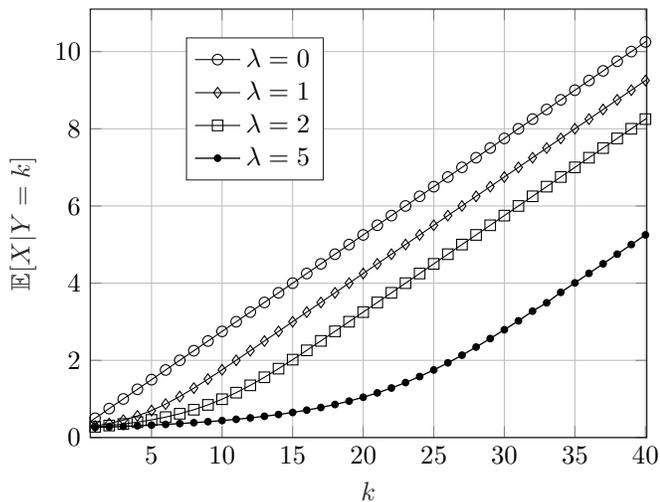
\begin{figure}[h!]  %[AD] Comment out for now
	\centering   
	%\input{FIG/EstimatorsExponentialDistribution.tex}%
\begin{tikzpicture}

\begin{axis}[%
width=7.4cm,
height=5.7cm,
at={(1.159in,0.77in)},
scale only axis,
xmin=0.656682027649771,
xmax=40.1,
xlabel style={font=\color{white!15!black}},
xlabel={$k$},
ymin=0,
ymax=11.1,
ylabel style={font=\color{white!15!black}},
ylabel={$\E[X|Y=k]$},
axis background/.style={fill=white},
xmajorgrids,
ymajorgrids,
legend style={at={(0.172,0.6)}, anchor=south west, legend cell align=left, align=left, draw=white!15!black}
]
\addplot [color=black, mark=o, mark options={solid, black}]
  table[row sep=crcr]{%
0	0.25\\
1	0.5\\
2	0.75\\
3	1\\
4	1.25\\
5	1.5\\
6	1.75\\
7	2\\
8	2.25\\
9	2.5\\
10	2.75\\
11	3\\
12	3.25\\
13	3.5\\
14	3.75\\
15	4\\
16	4.25\\
17	4.5\\
18	4.75\\
19	5\\
20	5.25\\
21	5.5\\
22	5.75\\
23	6\\
24	6.25\\
25	6.5\\
26	6.75\\
27	7\\
28	7.25\\
29	7.5\\
30	7.75\\
31	8\\
32	8.25\\
33	8.5\\
34	8.75\\
35	9\\
36	9.25\\
37	9.5\\
38	9.75\\
39	10\\
40	10.25\\
41	10.5\\
42	10.75\\
43	11\\
44	11.25\\
45	11.5\\
46	11.75\\
47	12\\
48	12.25\\
49	12.5\\
50	12.75\\
51	13\\
52	13.25\\
53	13.5\\
54	13.75\\
55	14\\
56	14.25\\
57	14.5\\
58	14.75\\
59	15\\
};
\addlegendentry{$\lambda=0$}

\addplot [color=black,  mark=diamond, mark size=2pt, mark options={solid, black}]
  table[row sep=crcr]{%
0	0.25\\
1	0.3\\
2	0.365384615384614\\
3	0.450704225352117\\
4	0.560679611650478\\
5	0.699066874027976\\
6	0.867162471395824\\
7	1.06274894295036\\
8	1.28042005822815\\
9	1.51333967313657\\
10	1.75530754887486\\
11	2.00192630005025\\
12	2.25064168854068\\
13	2.50019740249006\\
14	2.75005640002491\\
15	3.00001499019499\\
16	3.25000363968239\\
17	3.50000216420064\\
18	3.74999877385262\\
19	4.0000002621197\\
20	4.25000005518309\\
21	4.50000001103662\\
22	4.75000000210221\\
23	5.00000000038222\\
24	5.25000000006647\\
25	5.50000000001108\\
26	5.75000000000177\\
27	6.00000000000027\\
28	6.25000000000004\\
29	6.50000000000001\\
30	6.75\\
31	7\\
32	7.25\\
33	7.5\\
34	7.75\\
35	8\\
36	8.25\\
37	8.5\\
38	8.75\\
39	9\\
40	9.25\\
41	9.5\\
42	9.75\\
43	10\\
44	10.25\\
45	10.5\\
46	10.75\\
47	11\\
48	11.25\\
49	11.5\\
50	11.75\\
51	12\\
52	12.25\\
53	12.5\\
54	12.75\\
55	13\\
56	13.25\\
57	13.5\\
58	13.75\\
59	14\\
};
\addlegendentry{$\lambda=1$}

\addplot [color=black,  mark=square,  mark size=2pt, mark options={solid, black}]
  table[row sep=crcr]{%
0	0.25\\
1	0.277777777777777\\
2	0.310975609756097\\
3	0.350923482849605\\
4	0.399270482603816\\
5	0.458016606244889\\
6	0.529503770487892\\
7	0.616329399926177\\
8	0.721140522247445\\
9	0.846281655353462\\
10	0.993322128504657\\
11	1.16257664098021\\
12	1.35281277539743\\
13	1.56132926734424\\
14	1.78444178471602\\
15	2.01820177922015\\
16	2.25905964659576\\
17	2.50425438672973\\
18	2.75188902355175\\
19	3.00079420035606\\
20	3.25031847529539\\
21	3.5001374695329\\
22	3.74996857025491\\
23	4.00005617290663\\
24	4.25001957522359\\
25	4.50000653640201\\
26	4.75000209499854\\
27	5.00000064556972\\
28	5.25000019154265\\
29	5.50000005479636\\
30	5.75000001513423\\
31	6.00000000404028\\
32	6.25000000104374\\
33	6.50000000026119\\
34	6.75000000006338\\
35	7.00000000001492\\
36	7.25000000000342\\
37	7.50000000000076\\
38	7.75000000000016\\
39	8.00000000000004\\
40	8.25000000000001\\
41	8.5\\
42	8.75\\
43	9\\
44	9.25\\
45	9.5\\
46	9.75\\
47	10\\
48	10.25\\
49	10.5\\
50	10.75\\
51	11\\
52	11.25\\
53	11.5\\
54	11.75\\
55	12\\
56	12.25\\
57	12.5\\
58	12.75\\
59	13\\
};
\addlegendentry{$\lambda=2$}

\addplot [color=black, line width=0.5pt, mark size=1.2pt, mark=*, mark options={solid, black}]
  table[row sep=crcr]{%
0	0.250000000000003\\
1	0.261904761904759\\
2	0.27488687782805\\
3	0.289084280506111\\
4	0.30465677330415\\
5	0.321789935316197\\
6	0.340699728632882\\
7	0.361637904282631\\
8	0.384898294416287\\
9	0.410824051836695\\
10	0.439815843160338\\
11	0.472340903362047\\
12	0.508942696573384\\
13	0.550250675123275\\
14	0.596989257904066\\
15	0.649984637087228\\
16	0.710167365998604\\
17	0.77856792316349\\
18	0.856301717178283\\
19	0.944539549201386\\
20	1.04445980863317\\
21	1.15718015139286\\
22	1.28366976379085\\
23	1.42464813975017\\
24	1.58048343341781\\
25	1.75110938871019\\
26	1.93597305845446\\
27	2.13408253318659\\
28	2.34389671062399\\
29	2.56399018767538\\
30	2.79335879425582\\
31	3.02357447335541\\
32	3.27463310512646\\
33	3.49129143357618\\
34	3.76203119847275\\
35	4.00709438150826\\
36	4.25406525801466\\
37	4.50226511117587\\
38	4.75122799715708\\
39	5.0006481823217\\
40	5.25033332930249\\
41	5.50016711088253\\
42	5.75008172602924\\
43	6.00003901212864\\
44	6.25001818740805\\
45	6.50000828536139\\
46	6.75000369019011\\
47	7.0000016076811\\
48	7.25000068544534\\
49	7.50000028613208\\
50	7.75000011699623\\
51	8.00000004687828\\
52	8.25000001841373\\
53	8.50000000709334\\
54	8.75000000268078\\
55	9.00000000099433\\
56	9.25000000036208\\
57	9.50000000012949\\
58	9.75000000004549\\
59	10.0000000000157\\
};
\addlegendentry{$\lambda=5$}

\end{axis}

\end{tikzpicture}%
		\caption{Examples of conditional expectations for $X$ distributed according to an exponential distribution with rate parameter $\alpha=3$.} \label{fig:ConditionalExpectationExponential}
\end{figure}

By using the expression of the output distribution in Lemma~\ref{lem:OutputExponential} and the TGR formula in \eqref{eq:COnditionalExpectation}, Fig.~\ref{fig:ConditionalExpectationExponential} shows examples of conditional mean estimators for various values of the dark current parameter $\lambda$.  Observe that in Fig.~\ref{fig:ConditionalExpectationExponential} the estimator is linear for $\lambda=0$ and non-linear for $\lambda>0$. 

\subsection{Quantitative Refinement of the Linearity Condition in Theorem~\ref{thm:LinearityConditions}} 
\label{sec:quantitativeRefinementOfLinearity}
In this section, we refine the statement of Theorem~\ref{thm:LinearityConditions}. Specifically, we show that if the conditional expectation is close to a linear function in a mean squared error sense then the input distribution must be close in the L\'evy metric to the  gamma distribution where the  L\'evy metric is defined next. 
\begin{definition} Let $P$ and $Q$ be two cumulative distribution functions. Then, the \emph{L\'evy metric} between   $P$ and $Q$ is defined as 
\begin{align}
\mathbb{L}(P,Q)  =& \inf  \left\{  h  \ge 0:   Q(x-h)-h \le P(x)   \right. \notag  \\
& \qquad \left.     \le  Q(x+h)+h,   \forall x \in \mathbb{R} \right\}. 
\end{align}
\end{definition} 
An important property of the L\'evy metric is that  convergence of distributions in the L\'evy metric is equivalent to  weak convergence of distributions as defined in Definition~\ref{def:WeakConvergence} \cite{dudley2002real}.

\begin{theorem} \label{thm:LevyDistanceAndLInearity}  Let   $Y=\mathcal{P}(U)$ where $U \sim P_U$ and suppose that 
\begin{align}
\E \left[  \left|  \E[U|Y]  -   ( c_1Y+c_2) \right|^2  \right] \le \epsilon, \label{eq:MSEdiff}
\end{align} 
for some $0<c_1<1$ and $c_2>0$. 
Then, for $U_{\gamma} \sim \mathsf{Gam} \left( \frac{1-c_1}{c_1}, \frac{c_2}{c_1} \right)$
\begin{align}
 \sup_{s \ge 0} \left|  \frac{\phi_U(s)- \phi_{U_{\gamma}}(s)}{s}  \right|  \le  \frac{ \sqrt{\epsilon} }{1-c_1},  \label{eq:CharFunctionBounds}
\end{align} 
where  $\phi_U(s)$ and  $\phi_{U_{\gamma}}(s) $ are the characteristic functions of $U$ and $U_{\gamma}$,  respectively. Consequently, 
\begin{align}
\frac{\mathbb{L}^2 \left(P_U , \mathsf{Gam} \left( \frac{1-c_1}{c_1}, \frac{c_2}{c_1} \right) \right)}{2} \le  \frac{ \sqrt{\epsilon} }{1-c_1}.  \label{eq:BoundOnChars}
\end{align}
\end{theorem} 
\begin{IEEEproof}
See Appendix~\ref{app:thm:LevyDistanceAndLInearity}.
\end{IEEEproof}

\begin{remark}\label{rem:StabilityInTheGaussianCase}
The proof of Theorem~\ref{thm:LevyDistanceAndLInearity} is inspired by  the Gaussian analog  shown in \cite[Lemma~4]{du2018strong}.  
\end{remark} 

The bound in \eqref{eq:BoundOnChars} can also  be related  to the  Kolmogorov metric \cite{DytsoAsilomar2019}.

\section{Conclusion and Outlook} 
This work has focused on studying properties of the conditional mean estimator  of a random variable in Poisson noise.  Specific emphasis has been placed on how the conditional expectation behaves as a function of  the scaling parameter $a$ and dark current parameter  $\lambda$ and a function of the input distribution  (prior distribution). 

With respect to the channel parameters $(a,\lambda)$, several identities in terms of derivatives have been established.  These derivative identities have also been used to show that the conditional expectation is  a monotone function of both the dark current parameter $\lambda$ and the channel observation $y$.  Another such identity  has proposed a notion of a score function of the output pmf  where the  gradient is taken with respect to the channel parameters $(a,\lambda)$ (note that   derivatives with respect to the output space are not defined in the Poisson case), and has shown that this score function has a natural connection to the conditional expectation via a Tweedie-like formula.    In fact, by contrasting with a Gaussian case, it has been argued that differentiating with respect to the channel parameters $(a,\lambda)$ is a natural substitute for differentiation with respect to the output space as the latter  cannot be performed in view of discreteness of the output space.    Moreover,  in the processes of cataloging  Poisson identities that have  Gaussian counterparts,  new identities for higher moments have been found for the Gaussian noise case.  We refer the reader to Table~\ref{table:Identities} for the summary of all the identities and their Gaussian counterparts.  

\begin{table*}[h!]
\caption{Summary of Identities and Results.} \label{table:Identities}
\center
\begin{tabular}{|p{3.5cm}|  p{6.7cm}|  p{6.6cm}| }
\hline
Identity &  Poisson Channel $Y=\mathcal{P}(aX+\lambda)$ & Gaussian Channel $Y_\mathsf{G}=a V+N, \, N \sim \mathcal{N}(0,\sigma^2)$ \\
\hline   
Natural Transform and the Output Distribution  & Laplace (see Theorem~\ref{thm:LaplaceConnection}) & Weierstrass  (see Remark~\ref{rem:WeiestrssTransform})  \\ 
\hline
Derivative \& Difference   &  $ \begin{array}{l} \frac{\rm d}{ {\rm d}  a } P_{Y|X}(y|x)=    x  \frac{\rm d}{ {\rm d}  \lambda } P_{Y|X}(y|x) \\
=x \left(P_{Y|X}(y-1|x)-P_{Y|X}(y|x)\right) \end{array}  $ (see Lemma~\ref{thm:DerivativesOuputPMF})
  & $\frac{{\rm d}}{ {\rm d} a} f_{Y_{\mathsf{G}}|V}(y|v)= - \frac{v}{\sigma^2} \, \frac{{\rm d}}{ {\rm d} y} f_{Y_{\mathsf{G}}|V}(y|v)$ (see Remark~\ref{rem:RelationToDifferenceOperators})    \\ 
\hline 
Conditional Expectation \& Output Distribution &    $\E[X|Y=y]=  \frac{1}{a}  \frac{ (y+1) P_Y(y+1)}{P_Y(y)}-\frac{\lambda}{a}$ (see Lemma~\ref{lem:FormOFConditionalExpecttion})  &    $a \E[V|Y_\mathsf{G}=y]= y + \sigma^2  \frac{f_{Y_\mathsf{G}}'(y)}{f_{Y_\mathsf{G}}(y)} $ (see \eqref{eq:TweediesFormulaGaussian}) \\
\hline 
Higher Order Conditional Moments  &   $\E \left[ ( aX +\lambda)^k |Y =y \right]= \prod_{i=0}^{k-1}  \E \left[  aX +\lambda |Y =y+i \right]$   (see Lemma~\ref{lem:HigherMoments})&  $\begin{array}{l} \E[(aV)^k | Y_{\mathsf{G}}=y]   \\
=  \sigma^{2k} \eu^{-\frac{1}{\sigma^2} \int_0^y  \E[ a V | Y_{\mathsf{G}}=t] {\rm d} t}  \frac{{\rm d}^k}{ {\rm d} y^k }   \eu^{  \frac{1}{\sigma^2} \int_0^y  \E[ aV | Y_{\mathsf{G}}=t] {\rm d} t}\end{array} $ (see Remark~\ref{rem:GaussianHigherMoments})  \\
\hline
MMSE and Fisher Information  &    $\mmse(X|Y)= \frac{ a\E \left[  X  \right] +\lambda  - J^{\mathsf{Po}}(Y)}{a^2}$  (see Theorem~\ref{thm:BrownsIdentityPoisson}) &  $\mmse(X|Y_\mathsf{G})=  \frac{ \sigma^2-  \sigma^4 J(Y_\mathsf{G})}{a^2}$  (see Remark~\ref{rem:BrownGaussian}) \\
\hline 
Conditional Expectation \& Conditional Variance  &    ${\bf v} \boldsymbol{\cdot} \nabla_{{\bf v}} \E \left[X|Y=y \right]=  -    a \mathbb{V}(X|Y=y)$ (see Theorem~\ref{thm:MonotonictyOfConditionalExp})  &  $  \sigma^2 \frac{\rm d}{ {\rm d}y} \E[ a V|Y_\mathsf{G}=y]=  \mathbb{V}( aV|Y_\mathsf{G}=y)$  (see Remark~\ref{rem:HiherOrderDerivatives}) \\
\hline 
Uniqueness of the Conditional Expectation &   (see Theorem~\ref{thm:UniquenssOfConditionalExpectation})   &   (see Remark~\ref{rem:GaussianUniqueness}) \\ 
\hline 
Bounds on the Conditional Expectation &   $\E[ U| Y=y] = O(y)  $  (see Theorem~\ref{thm:AssimptoticUppoerBoun})  \newline $\E[ U| Y=y] = O(y \log(y))  $  (see Theorem~\ref{Thm:BoundOnConditionalExpectation})  &   $ | \E \left[V | Y_{\mathsf{G}}=y \right] | =O(|y|)$ (see Remark~\ref{rem:BoundOnConditionalExpectation}) \\ 
\hline 
Linearity of the Conditional Expectation &  iff  $X$ is according to Gamma and $\lambda=0$ (see Theorem~\ref{thm:LinearityConditions})   &  iif $V$ is according to Gaussian \\ 
\hline 
Stability of Linear Estimators  &  (see Theorem~\ref{thm:LevyDistanceAndLInearity})   &  (see Remark~\ref{rem:StabilityInTheGaussianCase}) \\ 
\hline 
\end{tabular}
\end{table*}

With respect to the input distribution, several new results have been shown. For instance, it has been shown that the conditional expectation can be written as a ratio of derivatives of the Laplace transform the input distribution.  This identity has been used to compute conditional expectations for a few commonly occurring   input distributions; see Table~\ref{table:CEexamples}.  Importantly,  it has been shown that the conditional expectation uniquely determines the input distribution  (i.e., the conditional expectation is a bijective operator of the input distribution).   Several consequences of this uniqueness have been discussed including the  uniqueness of least favorable distributions and  of minimax estimators.    Moreover, it has shown that the conditional mean estimator  is a linear function if and only if the dark current parameter is zero and the input distribution is a gamma distribution.  Furthermore, a quantitative refinement of the equality condition has been given by showing that if the conditional expectation is close to a linear function in $L_2$ distance, then the input distribution must be close to the gamma distribution in the L\'evy metric.  

We  conclude  the  paper  by  mentioning  a  few  interesting future directions:
\begin{itemize}[leftmargin=*]
\item It would be interesting to see to what extent the results of this paper can be generalized to the vector Poisson model. On the one hand,  results such as Lemma~\ref{lem:HigherMoments}, Theorem~\ref{thm:MonotonictyOfConditionalExp} and Theorem~\ref{thm:AssimptoticUppoerBoun} appear to have immediate extensions. On the other hand, results in Theorem~\ref{thm:LaplaceConnection} and Theorem~\ref{thm:LevyDistanceAndLInearity} might require more work. The interested reader is referred to \cite{dytso2020VectorPoisson}  for  preliminary results on these extensions.  
\item  Building on the uniqueness property of the conditional expectation it is  likely possible to show that the Bayesian risk  defined through the  Bregman  divergence  natural\footnote{Every distribution that is a member of an exponential family has a corresponding Bregman divergence, which is termed natural \cite{banerjee2005clustering}. } for the Poisson channel \cite{banerjee2005optimality}, that is 
\begin{align}
R&=\E \left[ \ell_{\mathcal{P}} (  X; \E[X|Y])  \right], \label{eq:PoissonBregman}\\
&\text{ where }  \ell_{\mathcal{P}} (  u; v)= u\log \frac{u}{v}-(u-v), \, v, u \ge 0,
\end{align}
is a \emph{strictly} concave function in the input distribution.  This together with the identities in \cite{guo2008mutualPoisson,atar2012mutual,wang2014bregman}, where it was shown that the mutual information between the  input $X$ and  the output $Y=\mathcal{P}(X)$  can be written as an integral of \eqref{eq:PoissonBregman},  would also imply that the mutual information is a \emph{strictly} concave function of the input distribution; and 
\item It would be interesting to understand how the optimal risk in \eqref{eq:PoissonBregman} compares to the linear risk, that is
\begin{align}
R_L&=\E \left[ \ell_{\mathcal{P}} (  X; c_1Y+c_2)  \right],
\end{align} 
where $c_1$ and $c_2$  are some constants.  Specifically, it would be interesting to study  the following limits: 
\begin{align}
&\lim_{a \to \infty}  \frac{R}{R_L} \label{eq:LImitA to Innfty},\\
&\lim_{a \to 0}  \frac{R}{R_L}  \label{eq:LImitA to 0},\\
& \lim_{\lambda \to \infty}  \frac{R}{R_L};  \label{eq:LImitCurrent to Infinity}
\end{align}
the limit in \eqref{eq:LImitA to Innfty}  focuses on the optimality of linear estimators   in a low noise regime  (a similar study for the Gaussian noise  was undertaken  in \cite{mmseDim}), and the limits in  \eqref{eq:LImitA to 0} and \eqref{eq:LImitCurrent to Infinity} focus on the optimality of the linear estimator in a low  signal regime (a similar study for the Gaussian noise was undertaken in \cite{GuoMMSEprop}). 
\end{itemize}

\begin{appendices}

\section{Proof of Theorem~\ref{thm:LaplaceConnection}} 
\label{app:thm:LaplaceConnection}

Let $U=aX+\lambda$. Then,  using the definition of the Laplace transform the output distribution can be written as 
\begin{align}
P_{Y}(y; P_{X})&=  \frac{1}{y!} \E\left[  U^y \eu^{-U} \right]\\
&= \frac{(-1)^y}{y!} \mathcal{L}_U^{(y)}(t) \Big |_{t=1}.  \label{eq:DerivativeLaplaceConnectionTOouputDistirbuiton}
\end{align}
Next, using the scaling and shifting properties of the Laplace transform we have that
\begin{align}
\mathcal{L}_U^{(y)}(t) \Big |_{t=1}&=\mathcal{L}_{aX+\lambda}^{(y)}(t) \Big |_{t=1}\\
&= \frac{{\rm d}^y}{{\rm d} t^y}\mathcal{L}_{X}(at)  \eu^{-t\lambda} \Big |_{t=1}\\
&=   \sum_{i=0}^y { y \choose i} a^{y-i}   \mathcal{L}_{X}^{(y-i)}(at)  (- \lambda)^{i}   \eu^{-\lambda t}   \Big |_{t=1} \label{eq:genProdRule}\\
&=   \sum_{i=0}^y { y \choose i} a^{y-i}   \mathcal{L}_{X}^{(y-i)}(a)  (- \lambda)^{i}   \eu^{-\lambda } ,
\end{align} 
where in \eqref{eq:genProdRule} we have used generalized product rule.  
This concludes the proof of \eqref{eq:LaplaceAndOutput}. 

To show \eqref{eq:ConnectionToLaplaceTransform}   observe that 
\begin{align}
\E \left[  U^y \eu^{-U} \right] = (-1)^y \frac{{\rm d}^y}{{\rm d} t^y}\mathcal{L}_U(t) \Big |_{t=1} .
\end{align}  
Therefore, we can write $\mathcal{L}_U(t)$ as a Taylor series around $t=1$ as follows:
\begin{align}
\mathcal{L}_U(t)&= \sum_{y=0}^\infty\frac{1}{y!} \frac{{\rm d}^y}{ {\rm d} u^y}\mathcal{L}_U(u) |_{u=1} (t-1)^y \\\
&= \sum_{y=0}^\infty   (-1)^y P_{Y}(y;P_X) (t-1)^y, \label{eq:PowerSeries}
\end{align}
where in the last step we have used  \eqref{eq:DerivativeLaplaceConnectionTOouputDistirbuiton}.  Therefore, it remains to find the region of convergence of \eqref{eq:PowerSeries}.  Next, by the root test for the convergence of power series we have that 
\begin{align}
\frac{1}{r}&= \limsup_{k \to \infty}    \left|  P_{Y}(k;P_X) \right|^{\frac{1}{k}} \le 1, \label{eq:RadiusOfConvergence}
\end{align}
where in the last  step we have used that $  P_{Y}(\, \cdot \,;P_X) \le 1$. 
From  \eqref{eq:RadiusOfConvergence} we have that the series in \eqref{eq:PowerSeries} converges  on the interval $|t-1|<1$.  
This concludes the proof.

\section{Proof of Lemma~\ref{thm:DerivativesOuputPMF}}
\label{app:thm:DerivativesOuputPMF}

In what follows we use the convention that $P_{Y|X}(-1|x)=P_Y(-1)=0$.  Proof of the expression in \eqref{eq:ChannelDerivatives} follows by inspection. 

To show \eqref{eq:GradientTypeConnection} observe that 
\begin{align}
&a \frac{{ \rm d} }{{ \rm d} a } P_{Y}(y)  \notag\\
&=y\E \left[a X \frac{(aX+\lambda)^{y-1}}{y!} \eu^{-(aX+\lambda)} \right]  \notag\\
 & - \E \left[aX  \frac{(aX+\lambda)^y}{y!} \eu^{-(aX+\lambda)} \right] \label{eq:ExchangeDiffAndExpect} \\
&=y\E \left[ \frac{(aX+\lambda)^{y}}{y!} \eu^{-(aX+\lambda)} \right]  \notag\\
&- \lambda  y   \E \left[ \frac{(aX+\lambda)^{y-1}}{y!} \eu^{-(aX+\lambda)} \right]  \notag\\
&  - \E \left[  \frac{(aX+\lambda)^{y+1}}{y!} \eu^{-(aX+\lambda)} \right] 
+ \lambda \E \left[  \frac{(aX+\lambda)^{y}}{y!} \eu^{-(aX+\lambda)} \right] \\
&=y  P_Y(y)-(y+1) P_Y(y+1)  + \lambda \left(    P_Y(y)- P_Y(y-1)  \right), 
\label{eq:DerivativeScaling}
\end{align}
the exchange of differentiation and expectation in \eqref{eq:ExchangeDiffAndExpect} follows from a simple application  of the dominated convergence theorem. 

Next,  we find the derivative with respect to $\lambda$ 
\begin{align}
 \frac{{ \rm d} }{{ \rm d} \lambda  } P_{Y}(y)  
 &=-  \E \left[  \frac{(aX+\lambda)^y}{y!} \eu^{-(aX+\lambda)} \right]   \notag\\
 &+ y  \E \left[  \frac{(aX+\lambda)^{y-1}}{y!} \eu^{-(aX+\lambda)} \right] \\
 &= -   P_Y(y) +   P_Y(y-1).
 \label{eq:DerivativeWithLambda}
\end{align} 
Finally,  combining \eqref{eq:DerivativeScaling} and \eqref{eq:DerivativeWithLambda} concludes the proof.

\section{Proof of Lemma~\ref{thm:TailBounds}} 
\label{app:thm:TailBounds}
The first upper bound in \eqref{eq:TailOverAllSitributions} follows by observing that the maximum of the function $ x \to  (ax+\lambda)^y \eu^{-(ax +\lambda)}$ occurs when $x=\frac{y-\lambda}{a} $. Therefore,
\begin{align}
P_Y(y) &=  \E \left[  P_{Y|X}(y|X) \right] \notag\\
&\le  \frac{1}{y!}  y^y \eu^{-y} \\
&\le   \frac{1}{\sqrt{2 \pi} \sqrt{y}} .\label{eq:TailOverAllSitributionsProof}
\end{align}
 The second bound in \eqref{eq:TailOverAllSitributionsProof} follows by using Stirling's lower bound $ y! \ge \sqrt{2 \pi}  y^{y+\frac{1}{2}} \eu^{-y}$.

We now show the lower bound in \eqref{eq:TailOverAllSitributions}. Let $U=aX+\lambda$.  Then,
\begin{align}
P_Y(y) &=  \frac{1}{y!} \E \left[ U^y \eu^{-U}  \right]\\
&=  \frac{1}{y!} \E \left[  \eu^{ y \log(U) -U }  \right]\\
& \ge    \frac{1}{y!}  \eu^{ y  \E[ \log(U)]-\E[U] }, 
\end{align} 
where in the last step we applied Jensen's inequality. 
This concludes the proof of the lower bound in \eqref{eq:TailOverAllSitributionsProof}. 

 \section{Proof of Theorem~\ref{thm:MonotonictyOfConditionalExp}}
\label{app:thm:MonotonictyOfConditionalExp}

In this proof, we use the convention that $P_{Y}(-1)=0$ and $\E[ U^k |Y= -1]=\infty$.
Fix some $y \ge 0$.  Then,
\begin{align}
& \frac{y!}{(y+k)!} \frac{ \rm d}{{\rm d} \lambda } \E[  U^k|Y=y] \notag\\
  &=   \frac{ \rm d}{{\rm d} \lambda }   \frac{ P_{Y}(y+k)}{P_{Y}(y)}  \label{eq:ApplicationOfHigherTGR} \\
 &=  \frac{ P_{Y}(y)  \frac{ \rm d}{{\rm d} \lambda }  P_{Y}(y+k)  - P_{Y}(y+k)  \frac{ \rm d}{{\rm d} \lambda } P_{Y}(y)     }{P_{Y}^2(y)} \\
 &=  \frac{ P_{Y}(y)  (P_{Y}(y+k-1) -   P_{Y}(y+k))    }{P_{Y}^2(y)} \notag\\ 
 &-   \frac{  P_{Y}(y+k) (P_{Y}(y-1) -   P_{Y}(y))    }{P_{Y}^2(y)} \label{eq:ApplyingDerivativeWithDarkCurrent}\\
 &=\frac{   P_{Y}(y+k-1)    }{P_{Y}(y)} -  \frac{  P_{Y}(y+k) P_{Y}(y-1)    }{P_{Y}^2(y)} \\
 &=    
  \frac{y!   \E[U^{k-1}|Y=y]}{ (y+k-1)!} -    \frac{ y!}{(y+k)!}    \frac{y   \E[U^k|Y=y]  }{  \E[U|Y=y-1] },  \label{eq:ApplicationOfHigherTGR_v2} 
\end{align}
where \eqref{eq:ApplicationOfHigherTGR} follows by using  the generalized TGR formula in \eqref{eq:Hihger Moments};  \eqref{eq:ApplyingDerivativeWithDarkCurrent} follows by using the derivative identity in \eqref{eq:DarkCurrentDerivative} where we use the convention that $P_{Y}(-1)=0$;  and  \eqref{eq:ApplicationOfHigherTGR_v2} follows by using  the generalized TGR formula in \eqref{eq:Hihger Moments} and the convention $\E[ U |Y= -1]=\infty$.  Dividing  \eqref{eq:ApplicationOfHigherTGR_v2}  by   $\frac{y!}{(y+k)!}$ leads to \eqref{eq:DerivativeOfHiherMoments}. 

For the case of $k=1$ \eqref{eq:ApplicationOfHigherTGR_v2} reduces to 
\begin{align}
 \frac{ \rm d}{{\rm d} \lambda } \E[  U|Y=y] =   (y+1)-  \frac{y \E[  U|Y=y] }{ \E[  U|Y=y-1]}.
\end{align}
Moreover,
\begin{align}
 a \frac{ \rm d}{{\rm d} \lambda } \E[  X|Y=y] =-  y      \frac{ \mathbb{V}(U|Y=y-1)}{  (\E \left[U|Y=y-1 \right])^2 } ,
\end{align}
where  in last step we have used the variance expression in \eqref{eq:COnditionalVariance}.

Next, we compute the derivative with respect to $a$ 
\begin{align}
& \frac{y!}{(y+k)!} \frac{ \rm d}{{\rm d} a } \E[  U^k|Y=y] \notag\\
  &=   \frac{ \rm d}{{\rm d} a }   \frac{ P_{Y}(y+k)}{P_{Y}(y)}   \label{eq:ApplicationOfHigherTGRv2} \\
  &=  \frac{ P_{Y}(y)  \frac{ \rm d}{{\rm d} a}  P_{Y}(y+k)  - P_{Y}(y+k)  \frac{ \rm d}{{\rm d} a } P_{Y}(y)     }{P_{Y}^2(y)} \\
  &=  \frac{  \frac{y+k}{a}  P_Y(y+k)- \frac{y+k+1}{a} P_Y(y+k+1)      }{P_{Y_{\lambda}}(y)}  \notag\\
   &+  \frac{   \frac{\lambda}{a}   \left(    P_Y(y+k)- P_Y(y+k-1)  \right)      }{P_{Y}(y)}  \notag\\
&- \frac{ P_{Y}(y+k) \left(   \frac{y}{a}  P_Y(y)- \frac{y+1}{a} P_Y(y+1)  \right)     }{P_{Y}^2(y)} \notag \\
&- \frac{ P_{Y}(y+k) \left(   \frac{\lambda}{a}   \left(    P_Y(y)- P_Y(y-1)  \right) \right)     }{P_{Y}^2(y)}  \label{eq:ApplicationDerivativePMFa}\\
&= 
\frac{y+k}{a}     \frac{y!  \E[  U^k|Y=y]}{(y+k)!}       - \frac{y+k+1}{a}  \frac{y!  \E[  U^{k+1}|Y=y]}{(y+k+1)!}     \notag\\
&+ \frac{\lambda}{a} \left(  \frac{y!  \E[  U^k|Y=y]  }{(y+k)!} -\frac{y!  \E[  U^{k-1}|Y=y]  }{(y+k-1)!}\right) \notag \\ 
&-    \frac{y!    \E[  U^k|Y=y]  \left(   \frac{y}{a}  - \frac{\E[U|Y=y]}{a}   + \frac{\lambda   \left(    1-  \frac{y}{ \E[  U|Y=y-1]} \right) }{a}  \right)  
  }{(y+k)!} \label{eq:ApplicationOfHigherTGR_v3} ,
     \end{align} 
  where  \eqref{eq:ApplicationOfHigherTGRv2}  follows by using  the generalized TGR formula in \eqref{eq:Hihger Moments};   and \eqref{eq:ApplicationDerivativePMFa}  follows by using the identity in \eqref{eq:DerivativeScaling}.  Next, dividing \eqref{eq:ApplicationOfHigherTGR_v3} by   $a \frac{y!}{(y+k)!} $ we have that
\begin{align}
& a \frac{ \rm d}{{\rm d} a } \E[  U^k|Y=y] \notag\\
&=k  \E[  U^k|Y=y]- \E[  U^{k+1}|Y=y]  \notag\\
&   + \E[  U^k|Y=y] \E[  U|Y=y] -\lambda (y+k)\E[  U^{k-1}|Y=y] \notag\\
&+ \frac{\lambda y \E[  U^k|Y=y] }{ \E[  U|Y=y-1]} .  \label{eq:DerivativeOfAFinal}
\end{align} 
Now, combining \eqref{eq:ApplicationOfHigherTGR_v2} and \eqref{eq:DerivativeOfAFinal} we have that 
\begin{align}
 & a \frac{ \rm d}{{\rm d} a } \E[  U^k|Y=y] + \lambda  \frac{ \rm d}{{\rm d} \lambda } \E[  U^k|Y=y]  \notag\\
&=k  \E[  U^k|Y=y]- \E[  U^{k+1}|Y=y]  \notag\\
&  + \E[  U^k|Y=y] \E[  U|Y=y],  \label{eq:CaseOfK===1}
\end{align} 
which concludes the proof of \eqref{eq:GradientOfCOnditionalExpectation}.

Setting $k=1$ the left  side of  \eqref{eq:CaseOfK===1} reduces to 
\begin{align}
&a \frac{ \rm d}{{\rm d} a } \E[  U|Y=y] + \lambda  \frac{ \rm d}{{\rm d} \lambda } \E[  U|Y=y]  \notag\\
&=  a \E[X|Y=y] + a^2  \frac{ \rm d}{{\rm d} a } \E[  X|Y=y] \notag\\
&+ a \lambda  \frac{ \rm d}{{\rm d} \lambda } \E[  X|Y=y] +\lambda ,  \label{eq:LeftHandSideOfk}
\end{align} 
and the right side of \eqref{eq:CaseOfK===1} reduces to 
\begin{align}
& \E[  U|Y=y]- \E[  U^{2}|Y=y]   + \E[  U|Y=y] \E[  U|Y=y]\notag\\
 &= \E[  U|Y=y] - \mathbb{V}(U|Y=y)\\
  &= a \E[ X|Y=y] +\lambda-  a^2\mathbb{V}(X|Y=y).   \label{eq:RightHandSideOfk}
\end{align} 

Combining \eqref{eq:LeftHandSideOfk} and \eqref{eq:RightHandSideOfk} we have that 
\begin{align}
&a \frac{ \rm d}{{\rm d} a } \E[  X|Y=y] + \lambda  \frac{ \rm d}{{\rm d} \lambda } \E[  X|Y=y] = -a \mathbb{V}(X|Y=y). 
\end{align}
This concludes the proof.

\section{Proof of Theorem~\ref{Thm:BoundOnConditionalExpectation}}
\label{app:Thm:BoundOnConditionalExpectation}

Before starting the proof, we will need the following definition and properties. 

\begin{definition} Consider the function
\begin{align}
w \eu^w=f(w)=x, \label{eq:LambertDef}
\end{align}
where $x$ and $w$ are real-valued.   The inverse of the function in \eqref{eq:LambertDef} is known as the \emph{Lambert W} function.    The  \emph{Lambert W} function  and solutions to \eqref{eq:LambertDef}  have the following properties:
\begin{itemize}[leftmargin=*]
\item    \eqref{eq:LambertDef} has a real-valued solutions only if   $x \ge - \frac{1}{\eu}$. Hence, the Lambert W function  is real-valued if  $x \ge - \frac{1}{\eu}$; and 
\item  \eqref{eq:LambertDef} has two solutions if   $ - \frac{1}{\eu}  <  x<0$ and a single solution for $x\ge 0$.   Hence, the Lambert W function has two real branches in the interval   $ - \frac{1}{\eu}  <  x<0$.  The two branches are denoted by $W_0$ (principle branch) and $W_{-1}$ (negative branch). 
\end{itemize} 
\end{definition}

We are now in position to proof the upper bound on the conditional expectation. Choose some $A>0$ and observe that 
\begin{align}  
&\E\left[ U \frac{P_{Y|U} (y|U) }{P_Y(y)} \right]  \notag\\
&=  \E\left[ U \frac{P_{Y|U} (y|U) }{P_Y(y)} 1_{\{ U P_{Y|U} (y|U) \le A  P_Y(y)   \}}\right] \notag\\
&+\E\left[ U \frac{P_{Y|U} (y|U) }{P_Y(y)} 1_{\{ U P_{Y|U} (y|U) > A P_Y(y)   \}}\right] \notag\\
&\le A + \E\left[ U \frac{P_{Y|U} (y|U) }{P_Y(y)} 1_{\{ U P_{Y|U} (y|U) > A P_Y(y)   \}}\right]. \label{eq:BoundConditionalDecomposition} 
\end{align} 

\begin{figure}[h!]
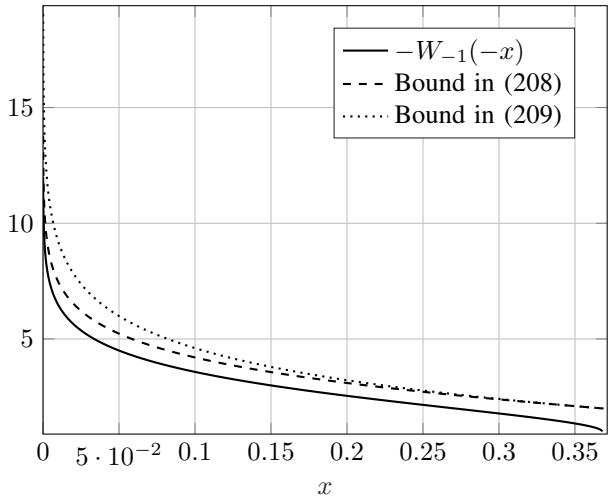
  %[AD] Comment out for now
	\centering   
\definecolor{mycolor1}{rgb}{0.00000,0.44700,0.74100}%

%
		\caption{Bounds on the Lambert W function.} \label{fig:BoundsLambert}
\end{figure}

Now note that the set in \eqref{eq:BoundConditionalDecomposition} can be re-written as follows:
\begin{align}
\left \{ x:  x P_{Y|U} (y|x) > A P_Y(y)  \right \} 
&= \left \{ x:  \frac{x^{y+1} e^{-x}}{y!} > A P_Y(y)   \right\}\\
&=\left \{ x:     g_l(y) < x  <  g_u(y)   \right\}, \label{eq:SetExpressionLambert}
\end{align} 
where
\begin{align}
 g_u(y) &=-(y+1) W_{-1} \left(- \frac{\left(A P_Y(y) y! \right)^{\frac{1}{y+1}}}{y+1}  \right) ,\label{eq:UpperLamberBound}\\
 g_l(y)&=-(y+1) W_{0} \left(- \frac{\left(A P_Y(y) y! \right)^{\frac{1}{y+1}}}{y+1}  \right) .\label{eq:LowerLamberBound}
\end{align}
 Moreover, the functions in \eqref{eq:UpperLamberBound}  and \eqref{eq:LowerLamberBound} are real-valued provided that
\begin{align}
A\le   \frac{1}{ P_Y(y) y!}  \left( \frac{y+1}{\eu}  \right)^{y+1}. \label{eq:COnditionFOrLInearity} 
\end{align}
By  combining   \eqref{eq:BoundConditionalDecomposition}  and  \eqref{eq:SetExpressionLambert} we have that
\begin{align}
\E[U|Y=y] & \le A +  g_u(y)   \\
&=A- (y+1) W_{-1} \left(- \frac{\left(A P_Y(y) y! \right)^{\frac{1}{y+1}}}{y+1}  \right).  \label{eq:BoundWithLambertFunction}
\end{align}
Next, we use the following bound on $W_{-1}$ shown in \cite{LamberWBOunds}: for $x \in [0,  \eu^{-1}]$
\begin{align}
-W_{-1}(-x) &\le \sqrt{ 2 \log \left (\frac{1}{x} \right)-1}+   \log \left(\frac{1}{x} \right)  \label{eq:WfunctionBoundFirst}\\
&\le  2  \log \left(\frac{1}{x} \right).  \label{eq:WfunctionBound}
\end{align}
Fig.~\ref{fig:BoundsLambert} compares the bounds in \eqref{eq:WfunctionBoundFirst} and \eqref{eq:WfunctionBound}. 

Collecting the bounds in  \eqref{eq:BoundWithLambertFunction} and \eqref{eq:WfunctionBound} we arrive at the following bound: 
\begin{align}
\E[U|Y=y] & \le A+ 2 (y+1)   \log \left( \frac{y+1}{\left(A P_Y(y) y! \right)^{\frac{1}{y+1}}}  \right).
\end{align}
Finally, by using the upper bound in \eqref{eq:TailOverAllSitributions},  it is not difficult to check that a choice of  $A=\eu^{-1}$ satisfies \eqref{eq:COnditionFOrLInearity}, and we arrive at
\begin{align}
&\frac{\E[U|Y=y]}{2}  \notag\\
& \le  \frac{1}{ 2 \eu}+1+ \log \left( \frac{1}{ P_Y(y) y! }  \right) +(y+1) \log(y+1)\\
&\le       \frac{1}{2\eu}+1 + \log \left( \frac{1}{ \eu^{ y  \E[\log(U)] -\E[U]  } }   \right) +(y+1) \log(y+1)\label{eq:Uisng the lower boundOnOutputPDF}\\
&=       \frac{1}{2\eu}+1 - y  \E[\log(U)]+ \E[U]  +(y+1) \log(y+1),
\end{align}
where the  inequality in \eqref{eq:Uisng the lower boundOnOutputPDF} follows by applying the lower bound in \eqref{eq:TailOverAllSitributions}.  This concludes the proof.

\section{Proof of Theorem~\ref{thm:LinearityConditions}}
\label{app:thm:LinearityConditions}

We begin the proof with the following lemma that will be useful for other purposes. 
\begin{lem}  Let $Y=\mathcal{P}(U)$. Then,  for any $t >0$
\begin{align}
& \E \left[  \left(U -(c_1Y+c_2) \right) \eu^{-tY} \right]   \notag\\
&  =    -\left(c_1(s-1)+1 \right) \mathcal{L}_U'(s)-c_2\mathcal{L}_U(s),
\label{eq:OrthgonalityIdenity}
\end{align}
where $s=1-\eu^{-t}$. 
\end{lem}
\begin{IEEEproof}
To compute \eqref{eq:OrthgonalityIdenity} we have to compute the following terms: 
\begin{align} 
\E \left[  U \eu^{-tY} \right], \, \E \left[  Y \eu^{-tY} \right],  \text{ and }  \E \left[   \eu^{-tY} \right]. \label{eq:setUpTODiffEquation}
\end{align} 

Now  we rewrite each term in \eqref{eq:setUpTODiffEquation} in terms of $U$ only.  To that end let
\begin{align}
v(t)=\eu^{-t}-1=-s,
\end{align}
 in which case $e^{-t}=1-s$ and $v'(t)=s-1$.

Now recall that the Laplace transform of a Poisson random variable $W$ with parameter $\beta$ is given by
\begin{align}
\E \left[   \eu^{-tW} \right]= \eu^{\beta v(t)}, \label{eq:LaplaceTransformOfPoisson}
\end{align}

Now 
\begin{align}
 \E \left[   \eu^{-tY} \right]&= \E \left[  \E \left[  \eu^{-tY} |U  \right] \right]\\
&=  \E \left[  \eu^{U v(t)} \right]\\
&= \mathcal{L}_U(s), \label{eq:LaplaceOfOutput}
\end{align}
where we used the fact that $Y$ given $U=u$  has a Poisson distribution  with a parameter $u$ and the Laplace transform of a Poisson random variable  in \eqref{eq:LaplaceTransformOfPoisson}.  Moreover, using similar steps
\begin{align}
\E \left[U e^{-tY} \right]&=\E \left[U \E \left[ e^{-tY}|U \right] \right]\\
&=\E \left[U  e^{U v(t)} \right]\\
&=\E \left[U  e^{-sU} \right ]\\
&=-\mathcal{L}_U'(s).
  \label{eq:Term2}
\end{align}
Finally,  
\begin{align}
\E \left[  Y \eu^{-tY} \right]
&= - \frac{ {\rm d}}{{\rm d}t} \E \left[   \eu^{-tY} \right] \label{eq:beforeTheDerivative}\\
&= - \frac{{\rm d}}{{\rm d}t} \E \left[   \eu^{U v(t)} \right]\\
&= -  \E \left[   \eu^{U v(t)} U v'(t) \right]\\
&=  (s-1) \mathcal{L}_U'(s),
 \label{eq:Term3}
\end{align}
where in \eqref{eq:beforeTheDerivative} we have used \eqref{eq:LaplaceOfOutput}. 
Now combining \eqref{eq:setUpTODiffEquation},  \eqref{eq:LaplaceOfOutput}, \eqref{eq:Term2}  and \eqref{eq:Term3} concludes the proof. 
\end{IEEEproof} 

Let $U=aX+\lambda$ and suppose that $\E[X|Y]=b_1 Y+b_2$ for some $b_1$ and $b_2$.   Then, from \eqref{eq:COnditionalExpectation} we have that 
\begin{align}
\E[U |Y=y]=c_1 y+c_2,
\end{align} 
with
\begin{align}
c_1&=a b_1, \label{eq:c1Toab1}\\
c_2&=a b_2+\lambda.  \label{eq:c2Tob2}
\end{align}

Then, by the orthogonality principle 
\begin{align}
0=\E \left[  \left(U -(c_1Y+c_2) \right) \eu^{-tY} \right],\label{eq:ConsequenceOfOrthogonality}
\end{align}
which in view of \eqref{eq:OrthgonalityIdenity} is equivalent to 
\begin{align}
-\mathcal{L}_U'(s)= c_1   (s-1) \mathcal{L}_U'(s)+ c_2 \mathcal{L}_U(s). 
\end{align}
Therefore, the final differential equation is given by
\begin{align}
-\left(  c_1s-  c_1+1 \right) \mathcal{L}_U'(s)=  c_2 \mathcal{L}_U(s),
\end{align}
where the boundary condition is given by 
\begin{align}
\mathcal{L}_U(0)=1.
\end{align}
The solution to this first-oder linear ordinary differential equation   is unique and is given by 
\begin{align}
\mathcal{L}_U(s)=  \frac{1}{ \left( 1+\frac{c_1}{1-c_1} s \right)^{\frac{c_2}{c_1}}}. \label{eq:LaplaceTransformGammaSolution}
\end{align}
The function in \eqref{eq:LaplaceTransformGammaSolution} is  the Laplace transform of  $U \sim  \mathsf{Gam} \left(  \frac{1-c_1}{c_1}, \frac{c_2}{c_1} \right)$.  

Next, observe that $\lambda=0$. This follows from  the definition of $U=a X+\lambda$ and the assumption that $X \ge 0$.  Since $U$ is distributed according to a gamma distribution,  a strictly positive $\lambda$ would  violated the fact that $X$ is a non-negative random variable.

Therefore, using \eqref{eq:c1Toab1} and \eqref{eq:c2Tob2}  we have that   $aX=U \sim  \mathsf{Gam} \left(  \frac{1-ab_1}{ab_1}, \frac{b_2}{ b_1} \right)$  and $X \sim  \mathsf{Gam} \left(  \frac{1-ab_1}{b_1}, \frac{b_2}{a b_1} \right)$.  This concludes the proof.

\section{Proof of Theorem~\ref{thm:LevyDistanceAndLInearity} }
\label{app:thm:LevyDistanceAndLInearity}

The following bounds on the  difference of two characteristic functions will be useful. 
\begin{lem} Let $\phi_{U}$ be characteristic function of some non-negative random variable $U$. Then, for any $\alpha, \theta >0$ and $ \tau>0$
\begin{align}
  \frac{\left| \phi_U(\tau)  \hspace{-0.02cm} -  \hspace{-0.02cm} \left(1 -\frac{i \tau}{\alpha} \right)^{-\theta} \right|  }{ \tau} \hspace{-0.02cm}
\le  \hspace{-0.03cm} \sup_{t \in [0,\tau]}  \left| \left(\frac{ t+i\alpha}{\alpha} \right)   \hspace{-0.02cm} \phi_U'(t)  \hspace{-0.02cm} +  \hspace{-0.02cm}  \frac{ \theta}{ \alpha}    \phi_U(t)  \right| .  \label{eq:BoundOnDiffOFCharFunctions}
\end{align}
\end{lem}
\begin{IEEEproof}
First observe that 
\begin{align}
&\phi_U(\tau) \left(1 -\frac{i \tau}{\alpha} \right)^\theta-1  \notag\\
&= \int_0^\tau  \left(1 -\frac{i t}{\alpha} \right)^\theta  \phi_U'(t)-  \frac{i \theta}{ \alpha}  \left(1 -\frac{i t}{\alpha} \right)^{\theta-1}  \phi_U(t) {\rm d} t\\
%&= \int_0^\tau  \left(1 -\frac{i t}{\alpha} \right)^\theta  \phi_U'(t)-  \frac{i \theta}{ \alpha}  \left(1 -\frac{i t}{\alpha} \right)^{\theta-1}  \phi(t) {\rm d} t\\
%&= \int_0^\tau  \left(1 -\frac{i t}{\alpha} \right)^{\theta-1} \left( \left(1 -\frac{i t}{\alpha} \right)  \phi_U'(t)-  \frac{i \theta}{ \alpha}    \phi(t)  \right) {\rm d} t\\
%&= \int_0^\tau  \left(1 -\frac{i t}{\alpha} \right)^{\theta-1} \left( \left(\frac{\alpha-i t}{\alpha} \right)  \phi_U'(t)-  \frac{i \theta}{ \alpha}    \phi(t)  \right) {\rm d} t\\
&= -\int_0^\tau i  \left(1 -\frac{i t}{\alpha} \right)^{\theta-1} \left( \left(\frac{ t+i\alpha}{\alpha} \right)  \phi_U'(t)+  \frac{ \theta}{ \alpha}    \phi_U(t)  \right) {\rm d} t.  \label{eq:GammaINtegraRepresentation}
\end{align}

Next, using the integral representation in \eqref{eq:GammaINtegraRepresentation}  and modulus inequality,  we have the following bound:
\begin{align}
& \left | \phi_U(\tau) \left(1 -\frac{i \tau}{\alpha} \right)^\theta-1 \right|  \notag\\
%&= \left | \int_0^\tau i  \left(1 -\frac{i t}{\alpha} \right)^{\theta-1} \left( \left(\frac{ t+i\alpha}{\alpha} \right)  \phi_U'(t)+  \frac{ \theta}{ \alpha}    \phi(t)  \right) {\rm d} t \right|\\
& \le   \int_0^\tau \left |  \left(1 -\frac{i t}{\alpha} \right)^{\theta-1} \right| \left| \left(\frac{ t+i\alpha}{\alpha} \right)  \phi_U'(t)+  \frac{ \theta}{ \alpha}    \phi_U(t)  \right| {\rm d} t \\
& =   \int_0^\tau   \left(1 +\frac{ t^2}{\alpha^2} \right)^{ \frac{\theta-1}{2} }  \left| \left(\frac{ t+i\alpha}{\alpha} \right)  \phi_U'(t)+ \frac{ \theta}{ \alpha}    \phi_U(t)  \right| {\rm d} t \\
& \le   \sup_{t \in [0,\tau]}  \left| \left(\frac{ t+i\alpha}{\alpha} \right)  \phi_U'(t)+ \frac{ \theta}{ \alpha}    \phi_U(t)  \right|   \int_0^\tau   \left(1 +\frac{ t^2}{\alpha^2} \right)^{ \frac{\theta-1}{2} } {\rm d} t .  \label{eq:ModulusIntegralBOund}
\end{align} 

To conclude the proof observe that the difference between characteristic functions is given by 
\begin{align}
&\left| \phi_U(\tau) - \left(1 -\frac{i \tau}{\alpha} \right)^{-\theta} \right| \notag\\
&= \left| \left(1 -\frac{i \tau}{\alpha} \right)^{-\theta} \right| \left | \phi_U(\tau) \left(1 -\frac{i \tau}{\alpha} \right)^\theta-1 \right|\\
&=  \left(1 +\frac{ \tau^2}{\alpha^2} \right)^{- \frac{\theta}{2}} \left | \phi_U(\tau) \left(1 -\frac{i \tau}{\alpha} \right)^\theta-1 \right|\\
&\le \sup_{t \in [0,\tau]}  \left| \left(\frac{ t+i\alpha}{\alpha} \right)  \phi_U'(t)+ \frac{ \theta}{ \alpha}    \phi_U(t)  \right|  \frac{  \int_0^\tau   \left(1 +\frac{ t^2}{\alpha^2} \right)^{ \frac{\theta-1}{2} } {\rm d} t }{   \left(1 +\frac{ \tau^2}{\alpha^2} \right)^{ \frac{\theta}{2}} } \label{eq:ApplyingBoundsOnDifferenceANDone}\\
&\le \sup_{t \in [0,\tau]}  \left| \left(\frac{ t+i\alpha}{\alpha} \right)  \phi_U'(t)+ \frac{ \theta}{ \alpha}    \phi_U(t)  \right|   \tau,  \label{eq:LastBOundOnDifferenceOfCharFunction}
\end{align} 
where in \eqref{eq:ApplyingBoundsOnDifferenceANDone} we have used the bound in \eqref{eq:ModulusIntegralBOund}; and in \eqref{eq:LastBOundOnDifferenceOfCharFunction} we use the bound that for every $\theta>0$
\begin{align}
\frac{  \int_0^\tau   \left(1 +\frac{ t^2}{\alpha^2} \right)^{ \frac{\theta-1}{2} } {\rm d} t }{   \left(1 +\frac{ \tau^2}{\alpha^2} \right)^{ \frac{\theta}{2}} }\le \tau. 
\end{align}
This concludes the proof.  
\end{IEEEproof}

Another useful result will the following bound on the L\'evy distance \cite{bohman1961} and  \cite{bobkov2016proximity}.
\begin{lem} Let $P$ and $Q$ be two distribution functions with characteristic functions $\phi_P$  and $\phi_Q$, respectively. Then, 
\begin{align}
\frac{\mathbb{L}^2(P,Q)}{2} \le \sup_{t  \ge 0 }   \left|  \frac{\phi_P(t)-\phi_Q(t)}{t}  \right|.  \label{eq:LevyAndChar}
\end{align} 
\end{lem}

We now proceed with the proof of the bound. Our starting place is the following consequence of the orthogonality principle:   fix some $t \in \mathbb{R}$ 
\begin{align}
0
&=\E \left[ (U -\E[U|Y]) \eu^{itY} \right] \notag\\
&=\E \left[ \left(U - (c_1Y+c_2)+ (c_1Y+c_2)-\E[U|Y] \right) \eu^{itY} \right]. \label{eq:OrthgonalityAndChar}
\end{align}
The identity in \eqref{eq:OrthgonalityAndChar} implies that
\begin{align}
&\E \left[ \left( c_1Y+c_2-\E[U|Y] \right) \eu^{itY} \right]  \notag\\
&=\E \left[ \left(U - (c_1Y+c_2) \right) \eu^{itY} \right]\\
&=  \frac{1}{i}  \frac{\rm d}{{\rm d}s} \phi_U(s)-c_1 (s-i) \frac{\rm d}{{\rm d}s} \phi_U(s) -c_2 \phi_U(s) \label{eq:UsingIdentityFOrCharButLaplace}\\
&=  - \left( \left( i(1-c_1)+c_1 s \right) \frac{\rm d}{{\rm d}s} \phi_U(s) +c_2 \phi_U(s) \right), \label{eq:ConsequenceOfOrthogonalityToChar}
\end{align} 
where in \eqref{eq:UsingIdentityFOrCharButLaplace}  $s$ and $t$ are related via  $i s=\eu^{it}-1$ and follows from  identity in \eqref{eq:OrthgonalityIdenity} by replacing the Laplace transform with the characteristic function.

Now applying the Cauchy-Schwarz inequality to \eqref{eq:ConsequenceOfOrthogonalityToChar} we have that
\begin{align}
& \left | \left( i(1-c_1)+c_1 s \right) \frac{\rm d}{{\rm d}s} \phi_U(s) +c_2 \phi_U(s)   \right| \notag\\
&=  \left | \E \left[ \left(  c_1Y+c_2-\E[U|Y] \right) \eu^{itY} \right]\right| \\
&\le \E \left[ \left|  c_1Y+c_2-\E[U|Y] \right|  \left| \eu^{itY} \right| \right] \\
&\le \sqrt{\E \left[ \left( c_1Y+c_2-\E[U|Y] \right)^2 \right]  \E \left[ \left |  \eu^{i tY}  \right |^2\right] } \\
&= \sqrt{\E \left[ \left( c_1Y+c_2-\E[U|Y] \right)^2 \right]  } . \label{eq:DifferenceLowerBound}
\end{align}

Setting $\alpha= \frac{1-c_1}{c_1}$ and $\theta= \frac{c_2}{c_1}$ and combining the bounds  in \eqref{eq:BoundOnDiffOFCharFunctions}  and \eqref{eq:DifferenceLowerBound}  we have that for all $s >0 $ 
\begin{align}
&\frac{1}{ s}\left| \phi_U(s) - \left(1 -\frac{i s}{\alpha} \right)^{-\theta} \right|   \notag\\
&\le  \frac{\sqrt{\E \left[ \left( c_1Y+c_2-\E[U|Y] \right)^2 \right]  } }{1-c_1}.  \label{eq:FinalConnectionBetwenCharAndMMSE}
\end{align}

Now applying \eqref{eq:FinalConnectionBetwenCharAndMMSE} and the bound in  \eqref{eq:LevyAndChar} we have that
\begin{align}
\frac{\mathbb{L}^2 \left(P_U , \mathsf{Gam} \left( \frac{1-c_1}{c_1}, \frac{c_2}{c_1} \right) \right)}{2}  
  &\le   \frac{ \sqrt{\E \left[ \left( c_1Y+c_2-\E[U|Y] \right)^2 \right]  } }{(1-c_1)} \\
  &\le \frac{ \sqrt{ \epsilon } }{(1-c_1)}.
\end{align} 
This concludes the proof.

\end{appendices}

\bibliography{refs}
\bibliographystyle{IEEEtran}

\end{document}